\documentclass[preprint,prd,aps,showpacs,showkeys,nofootinbib]{revtex4}
\usepackage{amsmath}
\usepackage{graphicx}
\usepackage{hyperref}
\usepackage{makecell}
\usepackage{bm}
\usepackage{url}
\usepackage{float} 
\usepackage{subfigure}
\begin{document}

\title{Higgs boson decays $h\rightarrow MZ$ in the TNMSSM}

\author{Huai-Cong Hu$^{1}$\footnote{huaiconghu@163.com},
Zhao-Yang Zhang$^{1}$\footnote{1311274306@qq.com},
Ning-Yu Zhu$^{1}$\footnote{hnzhuny@163.com},
Hai-Xiang Chen$^{1}$\footnote{haixchen@hotmail.com}}

\affiliation{$^1$School of Physical Science and Technology, Guangxi University, Nanning, 530004, China}

\begin{abstract}
We study the SM-like Higgs boson decays $h\rightarrow MZ$ in the Triplet extended NMSSM (TNMSSM), where $M$ is a vector meson $(\rho,\omega,\phi,J/\Psi,\Upsilon)$. Compared to the minimal supersymmetric standard model (MSSM), the TNMSSM includes two new SU(2) triplets with hypercharge ±1 and a SM gauge singlet which are coupled to each other. The indirect contributions to the decays $h\rightarrow MZ$ are produced from the effective $h\gamma Z$ vertex, and they are more important than the direct contributions. The results of this work would encourage a detection on $h\rightarrow Z\gamma$ at the future high energy colliders for exploring new physics beyond the SM. 
\date{\today}	
	
\end{abstract}
\maketitle

\section{Introduction}
Since the Higgs was discovered by the ATLAS and CMS collaborations in 2012\cite{mh-ATLAS,mh-CMS}, many questions regarding its properties remain unanswered. According to the latest experimental data the measured mass of the Higgs boson is\cite{PDG}.
\begin{eqnarray}
	m_h = 125.25 \pm 0.17 ~{\rm GeV}\nonumber
\end{eqnarray}
As a new elementary particle, $h$ is largely consistent with the neutral Higgs boson predicted by the SM. However many questions have been raised that challenge the SM framework.

Weak scale supersymmetry (SUSY) is a promising extension of the Standard Model (SM): it naturally explains why the electroweak (EW) symmetry breaking scale is much smaller than the Planck scale, and solve the gauge hierarchy problem\cite{Susskind:1978ms,Weinberg:1975gm}. But, minimum supersymmetric SM (MSSM) cannot fully solve the hierarchy problem. And it has another problem named “the $\mu$ problem”. Hence, extensions of the MSSM have been proposed to solve these problems. For example the extension of the MSSM by adding a (SM) gauge singlet which is coupled to Higgs doublets (the NMSSM) has been proposed to solve the $\mu$ problem. 

But unfortunately the NMSSM with all the couplings being perturbative up to the GUT scale also does not really solve the little hierarchy problem\cite{Ellwanger:2006rm,Ananthanarayan:1995zr,Ananthanarayan:1996zv,Mason:2009iq}. If taking little hierarchy problem seriously, then one should consider another source of Higgs quadruple coupling, which will not decouple in the large $\tan\beta$ limit\cite{TNMSSM}. The model which extended of the MSSM by adding SU(2) triplets(TMSSM)\cite{Espinosa:1991gr,Espinosa:1992hp,Espinosa:1998re} possess such a Higgs quartic coupling naturally.

Combined the advantages of the NMSSM and the TNMSSM, they can solve each other's problems\cite{TNMSSM}. In the triplet extended NMSSM (TNMSSM), the singlet interactions do not play any important role in raising the physical Higgs mass: one rely on triplets instead in achieving this goal. So, one also do not face the usual little hierarchy problems of the NMSSM.

In our work, we study 125 GeV SM-like Higgs boson decay $h\rightarrow MZ$ in the framework of the TNMSSM, with $M$ representing the mesons $\rho$, $\omega$, $\phi$, $J/\psi$ and $\Upsilon$. The decay of $h\rightarrow MZ$ have been shown in\cite{hMZ,hMZ1,Zhao:hmz,hMZ2,hMZ4,hMZ5,ATLAS:hMZ}, via the effective vertex $hZ\gamma^*$. Subsequent transition to $\gamma^*\rightarrow M$. There is no $hZ\gamma$ coupling at tree level, but it can be contributed by loop diagram\cite{hMZ}. The first evidence for the process $h\rightarrow\gamma Z$ is presented by the ATLAS and CMS. The obseved signal strength at the $68\%$ confidence level is $\mu=2.2^{+1.0}_{-0.9}$ for ATLAS analysis, $\mu=2.4^{+1.0}_{-0.9}$ for the CMS analysis, and $\mu=2.2\pm0.7$ for their combination\cite{ATLAS:2023yqk}.  Due to the process $h\rightarrow\gamma Z$ had observed and the results are shifted from the SM. This presupposes the existence of a new physics, whose contribution to the process may be able to explain the deviations between observed decays and SM predictions, and thus the associated decay process deserves to be investigated. This coupling is important for probing new physics. In the TNMSSM, there are additional coupling of the Higgs boson to additional charged scalars and charged fermions. They contribute to the $hZ\gamma$ coupling by loop diagrams.

The paper is organized as follows. we briefly present the main ingredients of the TNMSSM in Sec. II. We give the Higgs boson decays $h\rightarrow Z\gamma$ and $h\rightarrow MZ$ formulas in Sec. III. We show the input parameters and numerical results in Sec. IV. In the last section we give the discussion and conclusion. Finally, some relate formulas are shown in Appendix.

\section{The TNMSSM}
Compared to the MSSM, the TNMSSM includes two new SU(2)$_L$ triplet superfields $T$, $\bar T$ with hypercharge ±1 and a SM gauge singlet superfield $\hat s$ which are coupled to each other.

The superpotential of the TNMSSM can be written as:
\begin{eqnarray}
	W=-Y_d\hat d\hat q\hat H_d-Y_e\hat e\hat l\hat H_d+\chi_d\hat H_d \hat T\hat H_d+\lambda\hat H_u\hat H_d \hat s+\chi_t\hat H_u\hat{\bar{T}}\hat H_u+\frac13\kappa\hat s^3+\Lambda_T\hat s\hat{\bar T}\hat T+Y_u\hat u\hat q\hat H_u\nonumber
\end{eqnarray}
Here the triplet superfields with hypercharge Y = ±1 are defined as:
\begin{eqnarray}
	T\equiv T^a\sigma ^a=
	\begin{pmatrix}
		T^+/\sqrt{2}&-T^{++}\\
		T^0&-T^+/\sqrt{2}
	\end{pmatrix}\nonumber\\
	\bar{T}\equiv \bar{T}^a\sigma^a=
	\begin{pmatrix}
		\bar{T}^-/\sqrt{2}&-\bar{T}^{0}\\
		\bar{T}^{--}&-\bar{T}^-/\sqrt{2}
	\end{pmatrix}\nonumber
\end{eqnarray}
The soft SUSY breaking terms are shown as the follow
\begin{eqnarray}
	-\mathcal{L}_{\rm soft}=&&\frac{1}{2}(M_1\lambda_1\lambda_1+M_2\lambda_2\lambda_2+M_3\lambda_3\lambda_3  +h.c.)+m_{Hu}^2|H_u|^2+m_{H_d}^2|H_d|^2+m_{\bar T}^2Tr(\bar T^\dagger\bar T)+
	\nonumber\\
	&&m_T^2Tr(T^\dagger T)+m_S^2|S|^2 +m_{\tilde{Q}}^2|\tilde{Q}|^2+m_{\tilde{u}}^2|\tilde{u}_R|^2+m_{\tilde{d}}^2|\tilde{d}_R|^2+m_{\tilde{l}}^2|\tilde{l}|^2+m_{\tilde{e}}^2|\tilde{e}_R|^2+\nonumber\\
	&&(T_{\Lambda T} tr(T\bar T)S+T_{\chi d}H_d\cdot  TH_d+T_{\chi_t}H_u\cdot\bar{T}H_u+\frac{T_{\kappa}}{3}S^3+T_{\lambda}H_u\cdot H_dS+\nonumber\\ &&T_dH_d\cdot \tilde{Q}d^*-T_uH_u\cdot \tilde{Q}u^*+T_eH_d\cdot Le^*+h.c.),\label{soft}
\end{eqnarray}
where the respective definitions of the products
between two $SU(2)_L$ doublets and between a $SU(2)_L$ doublets and a $SU(2)_L$ triplet are given as follows:
\begin{eqnarray}
	&&H_u\cdot H_d=H_u^+H_d^--H_u^0H_d^0,\nonumber\\
	&&H_u\cdot \bar{T}H_u=\sqrt{2}H_u^0H_u^+\bar{T}^--(H_u^0)^2\bar{T}^0-(H_u^+)^2\bar{T}^{--},\nonumber\\
	&&H_d\cdot TH_d=\sqrt{2}H_d^0H_d^-T^+-(H_d^0)^2T^0-(H_d^-)^2T^{++}.\nonumber
\end{eqnarray}

Once the electroweak symmetry is spontaneously broken, the neutral scalar fields can be define as
\begin{eqnarray}
	&&\langle H_u^0\rangle=\frac{v_u+\phi_u+i\sigma_u}{\sqrt{2}}, 
	\qquad \langle H_d^0\rangle=\frac{v_d+\phi_d+i\sigma_d}{\sqrt{2}},\nonumber\\
	&&\langle T^0\rangle=\frac{v_T+\phi_T+i\sigma_T}{\sqrt{2}},\qquad \langle\bar{T}^0\rangle=\frac{\bar{v}_T+\phi_{\bar{T}}+i\sigma_{\bar{T}}}{\sqrt{2}},\nonumber\\
	&&\langle S\rangle=\frac{v_s+\phi_s+i\sigma_s}{\sqrt{2}}.\nonumber
\end{eqnarray}
and we define the ratio $v_u$ to $v_d$ as $\tan\beta=\frac{v_u}{v_d}$ the ratio $v_T$ to $v_{\bar T}$ as $\tan\beta'=\frac{v_T}{v_{\bar T}} $.

Since we introduce a single state and two triplet states, we have five minimization equations, including the usual upper and lower Higgs. In general, the vacuum expectation value of the triplet states must be small to avoid large $\rho$-parameter corrections\cite{TNMSSM}.

In the basis $(H_d^-,H_u^{+,*},\bar T^-,T^{+,*})$ and $(H_d^{-,*},H_u^+,\bar T^{-,*},T^+)$, the definition of the mass squared matrix for charged Higgs is given by

\begin{eqnarray}
	m_{H^-}^2=\begin{pmatrix}
		m_{H_d^-,H_d^{-,*}}&m_{H_u^{+,*},H_d^{-,*}}^*&m_{\bar T^-,H_d^{-,*}}^*&m_{T^{+,*},H_d^{-,*}}^*\\
		m_{H_d^-,H_u^+}&m_{H_u^{+,*},H_u^+}&m_{\bar T^-,H_u^+}^*&m_{T^{+,*},H_u^+}^*\\
		m_{H_d^-,\bar T^{-,*}}&m_{H_u^{+,*}\bar T^{-,*}}&m_{\bar T^-\bar T^{-,*}}&m_{T^{+,*}\bar T^{-,*}}^*\\
		m_{H_d^-,T^+}&m_{H_u^{+,*}T^+}&m_{\bar T^-T^+}&m_{T^{+,*}T^+}
	\end{pmatrix}
\end{eqnarray}
where
\begin{eqnarray}
	m_{H_d^-,H_d^{-,*}}=&&\frac{1}{2}v_s^2|\lambda|^2+\frac{1}{8}\big[g_1^2(2v_{\bar T}^2-2v_T^2-v_u^2+v_d^2)+g_2^2(-2v_{\bar T}^2+2v_T^2+v_d^2+v_u^2)\big]+v_d^2|\chi_d|^2+m_{H_d}^2,\nonumber\\
	m_{H_d^-,H_u^+}=&&\frac{1}{2}\big[\lambda(-v_dv_u\lambda^*+v_s^2\kappa^*-v_Tv_{\bar T}\Lambda_T^*)+\sqrt{2}v_sT_{\lambda}\big]+\frac{1}{4}g_2^2v_dv_u,\nonumber\\	m_{H_u^{+,*},H_u^+}=&&\frac{1}{2}v_s^2|\lambda|^2+\frac{1}{8}\big[(g_1^2+g_2^2)v_u^2-(-g_2^2+g_1^2)(2v_{\bar T}^2-2v_T^2+v_d^2)\big]+v_u^2|\chi_t|^2+m_{H_u}^2,\nonumber\\
	m_{H_d^-,\bar T^{-,*}}=&&\frac{1}{2\sqrt{2}}g_2^2v_dv_{\bar T}-\frac{1}{\sqrt{2}}v_s(v_d\chi_d\Lambda_T^*+v_u\lambda\chi_t^*),\nonumber\\
	m_{H_u^{+,*}\bar T^{-,*}}=&&\frac{1}{2\sqrt{2}}g_2^2v_{\bar T}v_u+\frac{1}{\sqrt{2}}(-2v_{\bar T}v_u\chi_t+v_dv_s\lambda)\chi_t^*-v_uT_{\chi,t^*},\nonumber\\
	m_{\bar T^-\bar T^{-,*}}=&&\frac{1}{2}v_s^2|\Lambda_T|^2+\frac{1}{4}\big[2g_2^2v_{\bar T}^2+(g_1^22v_{\bar T}^2-2v_T^2-v_u^2+v_d^2)\big]+v_u^2 |\chi_t|^2+m_{\bar T}^2,\nonumber\\
	m_{H_d^-,T^+}=&&\frac{1}{2\sqrt{2}}g_2^2v_dv_T+\frac{1}{\sqrt{2}}v_sv_u\chi_d\lambda^*-v_d(\sqrt{2}v_T|\chi_d|^2+T_{\chi_d}),\nonumber\\
	m_{H_u^{+,*}T^+}=&&\frac{1}{2\sqrt{2}}g_2^2v_dv_u-\frac{1}{\sqrt{2}}v_s(\Lambda_T v_u\chi_t^*+v_d\chi_d\lambda^*),\nonumber\\
	m_{\bar T^-T^+}=&&\frac{1}{2}g_2^2v_Tv_{\bar T}+\frac{1}{2}\big[\Lambda_T(-v_dv_u\lambda^*+v_s^2\kappa^*-v_Tv_{\bar T}\Lambda_T^*)+\sqrt{2}v_sT_{\Lambda_T}\big],\nonumber\\
	m_{T^{+,*}T^+}=&&\frac{1}{2}v_s|\Lambda_T|^2+\frac{1}{4}\big[2g_2^2v_T^2+g_1(-2v_{\bar T}^2+2v_T^2-v_d^2+v_u^2)\big]+v_d^2|\chi_d|^2+m_T^2.\nonumber
\end{eqnarray}
This matrix is diagonalized by $Z^+$:
\begin{eqnarray}
	Z^+m_{H^-}^2Z^{+,\dagger}=m_{2,H^-}^{\rm dia}\nonumber
\end{eqnarray}
with
\begin{eqnarray}
	H_d^-=\sum_jZ_{j1}^+H_j^-,~~~H_u^+=\sum_jZ_{j2}^+H_j^+,~~~T^-=\sum_jZ_{j3}^+H_j^-,~~~T^+=\sum_jZ_{j4}^+H_j^+.\nonumber
\end{eqnarray}

The mass of the SM-like Higgs boson in the TNMSSM can be written as:
\begin{eqnarray}
	m_h=\sqrt{(m^0_{h_1})^2+\Delta m_h^2}
\end{eqnarray}
where $m^0_{h_1}$ is the lightest tree-level Higgs boson mass, and $\Delta m^2_h$ is the radiative correction. The two-loop leading-log radiative corrections can be given as:
\begin{eqnarray}
	&&\Delta m^2_h=\frac{3m_t^4}{4\pi^2v^2}\left[\left(\tilde{t}+\frac{1}{2}\tilde{X}_t\right)+\frac{1}{16\pi^2}\left(\frac{3m_t^2}{2v^2}-32\pi\alpha_3\right)\left(\tilde{t}^2+\tilde{X}_t\tilde{t}\right)\right]\\
	&&\tilde{t}=\log\frac{M^2_S}{m^2_t},\qquad \tilde{X}_t=\frac{2\tilde{A}_t^2}{M_S^2}\left(1-\frac{\tilde{A}_t^2}{12M_S^2}\right)
\end{eqnarray} 
where $\alpha_3$ is the running strong coupling constant, $M_S=\sqrt{m_{\tilde{t}_1}m_{\tilde{t}_2}}$ with $m_{\tilde{t}_{1,2}}$ are the stop masses. $\tilde{A}_t=A_t-\mu\cot\beta$ with $A_t=T_{u,33}/Y_{u,33}$.

\section{Analytical formula}
\begin{figure}[h]
	\setlength{\unitlength}{1mm}
	\centering
	\includegraphics[width=1.4in]{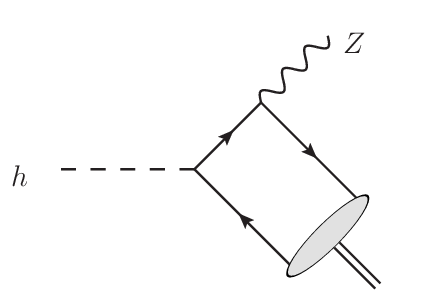}~\includegraphics[width=1.4in]{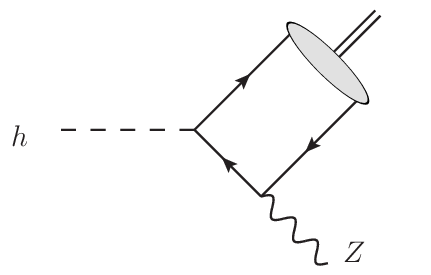}
	~\includegraphics[width=1.4in]{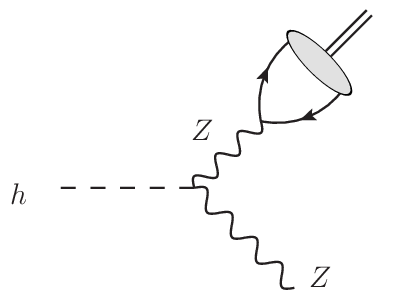}~\includegraphics[width=1.4in]{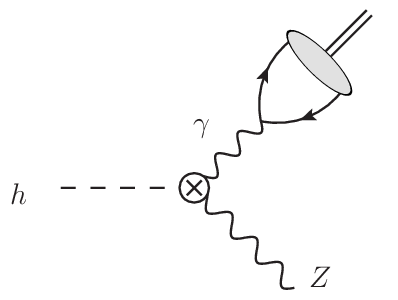}
	\caption[]{The diagrams contributing to the decay $h\rightarrow MZ$.
		The crossed circle in the last graph represents the effective vertex
		$h\rightarrow Z\gamma^*$ from the one loop diagrams.}\label{htoMVZ}
\end{figure}

In this section we discuss the Higgs boson decay processes $h\rightarrow Z\gamma$ and $h\rightarrow MZ$. The dominating Feynman diagrams for $h\rightarrow MZ$ are shown in Fig.\ref{htoMVZ}. The first two diagrams in Fig.\ref{htoMVZ} are the direct contributions, and the last two diagrams represent the indirect contributions. For the indirect contribution, there is a process $h\rightarrow Z\gamma^*\rightarrow MZ$, where $\gamma^*$ is off-shell and changes into the final state meson. 

There is no contribution to $hZ\gamma$ coupling at tree level in the TNMMSM, but it can be created by loop diagrams. The $h\rightarrow Z\gamma^*$ process can be used to probe
for New Physics. So we will focus on discussing $h\rightarrow Z\gamma^*$.  In the TNMSSM, the non-standard $h^0\gamma Z$ vertex should be taken into account. The effective Lagrangian for $h\gamma Z$ is written as:
\begin{eqnarray}
	\mathcal{L}_{eff}=\frac{\alpha}{4\pi \upsilon}\Big(\frac{2C_{\gamma Z}}{s_Wc_W}hF_{\mu\nu}Z^{\mu\nu}
	-\frac{2\tilde{C}_{\gamma Z}}{s_Wc_W}hF_{\mu\nu}\tilde{Z}^{\mu\nu}  \Big),\label{HZgammaE}
\end{eqnarray}
with $s_W=\sin\theta_W, c_W=\cos\theta_W$. The decay width of $h\rightarrow Z\gamma$ deduced by the effective Lagrangian defined in Eq.(\ref{HZgammaE}) is:
\begin{eqnarray}
	\Gamma(h\rightarrow Z\gamma)=\frac{\alpha^2m_{h^0}^3}{32\pi^3\upsilon^2s_W^2c_W^2}(1-\frac{m_Z^2}{m_{h^0}^2})^3(|C_{\gamma Z}|^2+|\tilde{C}_{\gamma Z}|^2).
\end{eqnarray}
The loop diagrams make additional contributions to $h\rightarrow MZ$ decays in new physics. And the decay width of $h\rightarrow MZ$ is given by:
\begin{eqnarray}
	&&\Gamma(h\rightarrow MZ)=\frac{m_{h^0}^3}{4\pi \upsilon^4}\,\lambda^{1/2}(1,r_Z,r_M)\,(1-r_Z-r_M)^2 \nonumber \\&&
	\quad\times \left[
	\big| F_\parallel^{MZ} \big|^2 + \frac{8r_M r_Z}{(1-r_Z-r_M)^2}
	\Big( \big| F_\perp^{MZ} \big|^2 + \big| \tilde F_\perp^{MZ} \big|^2 \Big) \right] ,\label{F}
\end{eqnarray}
with $\lambda(x,y,z)=(x-y-z)^2-4yz,r_Z=\frac{m_Z^2}{m_h^2}$ and $ r_M=\frac{m_M^2}{m_h^2}$, $m_M$ is the mass of vector meson. $ F_{\parallel}^{MZ}$ and $F_{\perp}^{MZ}$ represent the CP-even longitudinal and  transverse form factors, respectively $\tilde F_{\perp}^{MZ}$ represent the CP-odd transverse form factors. For the vector mesons considered in this work, the mass ratio $r_M$ is very small, but it can make significant contributions to the transverse polarization states. In order to obtain better results we keep them in our study.

In Eq.(\ref{F}), $F_{\parallel}^{MZ}, F_{\perp}^{MZ}$ and $\tilde F_{\perp}^{MZ}$ could be divided into direct and indirect parts. The indirect contributions are shown as follows:
\begin{eqnarray}
	&&F_{\parallel\, \rm indirect}^{MZ}= \frac{\kappa_Z}{1-r_M/r_Z} \sum_q f_M^q\,v_q
	+ C_{\gamma Z}\,\frac{\alpha_s (m_M)}{4\pi}\,\frac{4r_Z}{1-r_Z-r_M} \sum_q f_M^q\,Q_q\label{c1},\\
	&&F_{\perp\, \rm indirect}^{MZ}=\frac{\kappa_Z}{1-r_M/r_Z} \sum_q f_M^q\,v_q
	+ C_{\gamma Z}\,\frac{\alpha_s (m_M)}{4\pi}\,\frac{1-r_Z-r_M}{r_M} \sum_q f_M^q\,Q_q\label{c2},\\
	&&\tilde F_{\perp\, \rm indirect}^{MZ}=\tilde C_{\gamma Z}\,\frac{\alpha_s (m_M)}{4\pi}\,\frac{\lambda^{1/2}(1,r_Z,r_M) }{r_M} \sum_q f_M^q\,Q_q\label{c3},
\end{eqnarray}
where $v_q=\frac{T_3^q}{2}-Q_q\sin^2 \theta_W$ are the vector couplings of the Z boson to the quark $q$, $\kappa_Z$ is the ratio of the coupling of the SM-like Higgs boson to $Z$ boson to the corresponding SM value. $\alpha_s$ is the strong coupling constant. The flavor-specific decay constants $f_M^q$ are defined by
\begin{eqnarray}
	\langle M(k,\varepsilon)|\bar q\gamma^\mu q|0\rangle=-if_M^qm_M\varepsilon^{*\mu},~~~~~q=u,d,s\dots
\end{eqnarray}
The calculations can be simplified by the following relationship
\begin{eqnarray}
	\sum_q f_M^qQ_q=f_MQ_M, ~~~~~~\sum_q f_M^qv_q=f_Mv_M.
\end{eqnarray}
The vector meson decay constants $f_M, Q_M, v_M$ are shown in Table \ref{t1}.
\begin{table}[ht]
	\begin{tabular}{|c|c|c|c|c|c|}
		\hline
		Vector meson & $\omega$ & $\rho$ & $\phi$ & $J/\psi$ & $\Upsilon$ \\
		\hline
		$m_M/{\rm GeV}$ & 0.782 & 0.77 & 1.02 & 3.097 & 9.46 \\
		\hline
		$f_M/{\rm GeV}$ & 0.194 & 0.216 & 0.223 & 0.403 & 0.684 \\
		\hline
		$v_M$ &$ -\frac{\sin^2\theta_W}{3 \sqrt{2}}$ & $\frac{1}{\sqrt{2}} (\frac{1}{2}-\sin^2\theta_W)$ & $-\frac{1}{4}+\frac{\sin^2\theta_W}{3}$ & $\frac{1}{4}-\frac{2 \sin^2\theta_W}{3}$ & $-\frac{1}{4}+\frac{\sin^2\theta_W}{3}$  \\
		\hline
		$Q_M$ & $\frac{1}{3 \sqrt{2}}$ & $\frac{1}{\sqrt{2}}$ & $-\frac{1}{3}$ & $\frac{2}{3}$ & $-\frac{1}{3}$ \\
		\hline
		${f^{\perp}_M}/{f_M}={f^{q \perp}_M}/{f^{q}_M}$ & 0.71 & 0.72 & 0.76 & 0.91 & 1.09 \\
		\hline
	\end{tabular}
	\caption{The mesons decay constants $f_M$, $Q_M$, $v_M$ will be used in the numerical analysis, where $f_M^{\perp}$ and $f_M^{q\perp}$ represent the transverse decay constants and the flavor-specific transverse decay constants.}
	\label{t1}
\end{table}

The concrete forms of $C_{\gamma Z}$ and $\tilde C_{\gamma Z} $ in Eqs. (\ref{c1}-\ref{c3}) can be written as\cite{Zhao:hmz,cgz,hMZ2}
\begin{eqnarray}
	 &&C_{\gamma Z}= C_{\gamma Z}^{SM}+C_{\gamma Z}^{NP}, ~~~~~~~~~~~~~~   \tilde C_{\gamma Z}
	= \tilde C_{\gamma Z}^{SM}+\tilde C_{\gamma Z}^{NP}.\nonumber\\
	&&C_{\gamma Z}^{SM}= \sum_q \frac{2 N_c Q_q v_q}{3}\,A_f(\tau_q,r_Z)
	+ \sum_l \frac{2 Q_l v_l}{3}\,A_f(\tau_l,r_Z)
	- \frac{1}{2}\,A_W^{\gamma Z}(\tau_W,r_Z),\nonumber\\
	&&\tilde{C}_{\gamma Z}^{SM}=\sum_{q} \tilde{\kappa}_q N_c Q_q v_q B_f({\tau}_q,r_Z)+\sum_{l} \tilde{\kappa}_l Q_l v_l B_f({\tau}_l,r_Z),
\end{eqnarray}
where $\tau_i=\frac{4m_i^2}{m_h^2}$, $v_l$ are the vector couplings of the Z boson to the leptons are the vector couplings of the Z boson to the leptons and $Q_l$ represent the charge of leptons. $C_{\gamma Z}^{\rm SM}$ and $\tilde{C}_{\gamma Z}^{\rm SM}$ represent the SM contributions to $h\rightarrow Z\gamma$. $\tilde\kappa_q$ and $\tilde{\kappa}_l$ represent the effective Higgs couplings to the quarks and the leptons. $A_f,~ B_f$, and $A_W^{\gamma Z}$ are loop function could find in Refs. \cite{hMZ4,higgs decay1}. The numerical values of $C_{\gamma Z}^{\rm SM}$ and $\tilde C_{\gamma Z}^{\rm SM}$ are taken as : $C_{\gamma Z}^{\rm SM}\sim-2.395+0.001i,~\tilde C_{\gamma Z}^{\rm SM}\sim0$ in Ref. \cite{hMZ2}.

\begin{figure}
	\setlength{\unitlength}{1mm}
	\begin{minipage}[c]{0.8\textwidth}
		\centering
		\includegraphics[width=5.5in]{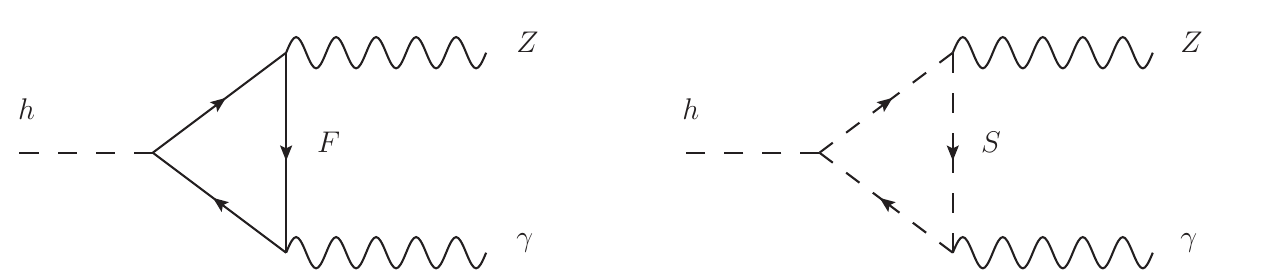}
	\end{minipage}%
	\caption[]{The one loop diagrams for $h\rightarrow \gamma Z$ in the TNMSSM, with $F=\chi^{\pm},\chi^{\pm\pm}$  denoting the charged fermions and $S=\tilde{u}_{i}^{+},\tilde{d}_{i}^{-},S_{\alpha}^{\pm},S_{\alpha}^{\pm\pm}$ denoting the squarks and charged scalars.}
	\label{OLDhzg}
\end{figure}

In the TNMSSM, the one loop diagrams contributing to $h\rightarrow\gamma Z$ are shown in Fig.\ref{OLDhzg}, where F represent the charged Fermions and S represent the charged scalars. The new contributions to $C_{\gamma Z}$ originate from the exchanged particles:charginos, sleptons, squarks, and charged Higgs.

The CP-odd coupling $\widetilde{C}_{\gamma Z}$ is 0 in the SM. In the TNMSSM, the $h\gamma Z$ coupling can be written as $\bar{F}_1 i(A+B\gamma^5)F_2h $, where $A$ is the CP-even part and $B$ is the CP-odd part\cite{hMZ2,Zhao:hmz}. For interaction $\bar{F}_1i(C^RP_R+C^LP_L)F_2h$ with $P_L=\frac{1-\gamma_5}{2}$ and $P_R=\frac{1+\gamma_5}{2}$, the CP-even part is $A=\frac{1}{2}(C^L+C^R)$ and the CP-odd part is $B=\frac{1}{2}(C^R-C^L)$. In the TNMSSM, 
\begin{eqnarray}
&&C^L_{hS^+S^-}=C^R_{hS^+S^-}, ~~C^L_{hS^{++}S^{--}}=C^R_{hS^{++}S^{--}},~~C^L_{h\tilde{f}\tilde{f}}=C^R_{h\tilde{f}\tilde{f}},\nonumber\\ 
&&C^L_{h\chi^+\chi^-}=C^R_{h\chi^+\chi^-},~~C^L_{h\chi^{++}\chi^{--}}=C^R_{h\chi^{++}\chi^{--}}\nonumber                   
\end{eqnarray}
and only $C^L_{Z\chi^+\chi^-}\neq C^R_{Z\chi^+\chi^-}$ or $C^L_{Z\chi^{++}\chi^{--}}\neq C^R_{Z\chi^{++}\chi^{--}}$. 
But $C^L_{Z\chi^+\chi^-}+C^R_{Z\chi^+\chi^-}\gg C^R_{Z\chi^+\chi^-}-C^L_{Z\chi^+\chi^-}$, $C^L_{Z\chi^{++}\chi^{--}}+C^R_{Z\chi^{++}\chi^{--}}\gg C^R_{Z\chi^{++}\chi^{--}}-C^L_{Z\chi^{++}\chi^{--}}$. So we can neglect the CP-odd coupling $\widetilde{C}^{NP}_{\gamma Z}$ in the TNMSSM. The expression of CP-even coupling $C^{NP}_{\gamma Z}$ in the TNMSSM is:
\begin{eqnarray}
	&&C_{\gamma Z}^{NP}=\frac{c_w}{2}\big[\sum_{S^{\pm} }\left(2c_w^2-1\right)g_{hS^+S^-}(m_Z^2/m_{S^{\pm}}^2)A_0[x_{S^\pm},\lambda_{S^{\pm}}]\nonumber\\	&&\quad\qquad+\sum_{S^{\pm\pm}}\left(2c_w^2-1\right)g_{hS^{++}S^{--}}(m_Z^2/m_{S^{\pm\pm}}^2)A_0[x_{S^{\pm\pm}},\lambda_{S^{\pm\pm}}]\nonumber\\
	&&\quad\qquad+\sum_{\tilde{f}}N_cQ_{\tilde{f}}\hat{v}_{\tilde{f}}g_{h\tilde{f}\tilde{f}}(m_Z^2/m_{\tilde{f}}^2)A_0[x_{\tilde{f}},\lambda_{\tilde{f}}]\nonumber\\
	&&\quad\qquad+\sum_{\chi^{\pm};m,n=L,R}g^m_{h\chi^{\pm}\chi^{\mp}}g^n_{Z\chi^{\pm}\chi^{\mp}}(2m_W/m_{\chi^{\pm}})A_{1/2}[x_{\chi^\pm},\lambda_{\chi^\pm}]\nonumber\\
	&&\quad\qquad+\sum_{\chi^{\pm\pm};m,n=L,R}g^m_{h\chi^{\pm\pm}\chi^{\mp\mp}}g^n_{Z\chi^{\pm\pm}\chi^{\mp\mp}}(2m_W/m_{\chi^{\pm\pm}})A_{1/2}[x_{\chi^\pm\pm},\lambda_{\chi^\pm\pm}]\big] 
\end{eqnarray}
where $x_i=4m_i^2/m_h^2$, $\lambda_i=4m_i^2/m_Z^2$, $\hat{v}_{\tilde{f}_1}=(T^f_3\cos^2\theta_f-Q_fs_w^2)/c_w$, $\hat{v}_{\tilde{f}_2}=(T^f_3\sin^2\theta_f-Q_fs_w^2)/c_w$, $\hat{v}_{\tilde{f}_1}$ and $\hat{v}_{\tilde{f}_2}$ represent up and down-quark sectors, $T_3^f$ is the weak isospin of fermion $f$, $\theta_f$ is the mixing angle of sfermins $\tilde{f}_{1,2}$. The function $A_0$, $A_{1/2}$ can be found at\cite{higgs decay1,higgs decay2}. The concrete expressions of couplings
\begin{eqnarray}
g_{hS^+S^-}=-\frac{v}{2m_Z^2}C_{hS^+S^-}^{L,R}&&,~g_{hS^{++}S^{--}}=-\frac{v}{2m_Z^2}C_{hS^{++}S^{--}}^{L,R}\nonumber\\
g_{h\tilde{f}\tilde{f}}=-\frac{v}{2m_Z^2}C_{h\tilde{f}\tilde{f}}^{L,R}&&,~g_{h\chi^{\pm}\chi^{\mp}}^{L,R}=-\frac{1}{e}C_{h\chi^{\pm}\chi^{\mp}}^{L,R}\nonumber\\
g_{Z\chi^{\pm}\chi^{\mp}}^{L,R}=-\frac{1}{e}C_{Z\chi^{\pm}\chi^{\mp}}^{L,R}&&,~g_{h\chi^{\pm\pm}\chi^{\mp\mp}}^{L,R}=-\frac{1}{e}C_{h\chi^{\pm\pm}\chi^{\mp\mp}}^{L,R}\nonumber\\ g_{Z\chi^{\pm\pm}\chi^{\mp\mp}}^{L,R}=-\frac{1}{e}C_{Z\chi^{\pm\pm}\chi^{\mp\mp}}^{L,R}&&.\nonumber
\end{eqnarray}

As discussed in Ref.\cite{QCD}, the QCD corrections to the process $h\rightarrow Z\gamma$ are around $0.1\%$ which medicates the QCD corrections can be neglected safely. In other words, we can safely neglect QCD corrections because they are very small.

Compared to the indirect contributions, the direct contributions are very different, and they can be calculated in a power series $(m_q/m_h)^2$ or $(\Lambda_{QCD}/m_h)^2$. For the transversely polarized vector meson, leading-twist projections provide direct contributions. We can get the direct contributions by the asymptotic function $\phi_M^{\perp}=6x(1-x)$ \cite{direct,direct1,direct2}

\begin{eqnarray}
	F_{\perp\rm direct}^{MZ}=\sum_qf_M^{q\perp}v_q\kappa_q\frac{3m_q}{2m_M}\frac{1-r_Z^2+2r_Z\ln r_Z}{(1-r_Z)^2}\\
	\tilde{F}_{\perp\rm direct}^{MZ}=\sum_qf_M^{q\perp}v_q\tilde\kappa_q\frac{3m_q}{2m_M}\frac{1-r_Z^2+2r_Z\ln r_Z}{(1-r_Z)^2}
\end{eqnarray}
In the calculations, it is found that the direct contribution is much smaller than the indirect contribution. Which indicates that the indirect contributions are more important than the direct contributions. The contributions for the decay width of $h\rightarrow MZ$ in SM are shown in Table \ref{t2}.
\begin{table}[ht]
	\begin{tabular}{| c | c | c | c | c | c |}
		\hline
		\quad & ${\rho}$ & $\omega$ & $\phi$ & J/$\psi$ & $\Upsilon$ \\
		\hline
		$F_{\parallel ind}^{MZ}$ & $\makecell[c]{0.0423\\+4.3\times10^{-4}C_{\gamma Z}^{SM}}$ & $\makecell[c]{-0.0102\\+1.3\times10^{-4}C_{\gamma Z}^{SM}}$ & $\makecell[c]{-0.0392\\-2.1\times10^{-4}C_{\gamma Z}^{SM}}$ & $\makecell[c]{0.041\\+7.5\times10^{-4}C_{\gamma Z}^{SM}}$ & $\makecell[c]{-0.115\\-6.1\times10^{-4}C_{\gamma Z}^{SM}}$ \\
		\hline
		$F_{\perp ind}^{MZ}$ & $\makecell[c]{0.042\\+1.181C_{\gamma Z}^{SM}}$ & $\makecell[c]{-0.01\\+0.343C_{\gamma Z}^{SM}}$ & $\makecell[c]{-0.039\\-0.327C_{\gamma Z}^{SM}}$ & $\makecell[c]{0.04\\+0.128C_{\gamma Z}^{SM}}$ & $\makecell[c]{-0.12\\-0.011C_{\gamma Z}^{SM}}$ \\
		\hline
		$F_{\perp direct}^{MZ}$ & $0.0037$ & $-0.00087$ & $-0.00257$ & $-0.00088$ & $-0.00080$ \\
		\hline
	\end{tabular}
	\caption{The contributions for the decay width of $h\rightarrow MZ$ in SM, with $C_{\gamma Z}^{SM}\simeq-2.43$.}
	\label{t2}
\end{table}

Normalized to the SM expectation, the signal strengths for the Higgs decay channels can be
quantified as
\begin{eqnarray}
	&&\mu^{ggF}_{MZ}=\frac{\sigma_{\rm NP}(ggF)Br_{\rm NP}(h\rightarrow MZ)}{\sigma_{\rm SM}(ggF)Br_{\rm SM}(h\rightarrow MZ)}\label{signal strengths1}\\
	&&\mu^{ggF}_{\gamma\gamma}=\frac{\sigma_{\rm NP}(ggF)Br_{\rm NP}(h\rightarrow \gamma\gamma)}{\sigma_{\rm SM}(ggF)Br_{\rm SM}(h\rightarrow \gamma\gamma)}\label{signal strengths2}
\end{eqnarray}
where ggF stands for gluon-gluon fusion. And Higgs production cross sections can be written as
\begin{eqnarray}
	\frac{\sigma_{\rm NP}(ggF)}{\sigma_{\rm SM}(ggF)}\approx\frac{\Gamma_{\rm NP}(h\rightarrow gg)}{\Gamma_{\rm SM}(h\rightarrow gg)}\label{signal strengths3}
\end{eqnarray}
Through Eqs.(\ref{signal strengths1}-\ref{signal strengths3}), the signal strengths for $h\rightarrow MZ$ and $h\rightarrow \gamma\gamma$ can be quantified as
\begin{eqnarray}
	&&\mu^{ggF}_{MZ}\approx\frac{\Gamma_{\rm NP}(h\rightarrow gg)\Gamma_{\rm NP}(h\rightarrow MZ)/\Gamma^h_{\rm NP}}{\Gamma_{\rm SM}(h\rightarrow gg)\Gamma_{\rm SM}(h\rightarrow MZ)/\Gamma^h_{\rm SM}}=\frac{\Gamma_{\rm NP}(h\rightarrow gg)\Gamma_{\rm NP}(h\rightarrow MZ)\Gamma^h_{\rm SM}}{\Gamma_{\rm SM}(h\rightarrow gg)\Gamma_{\rm SM}(h\rightarrow MZ)\Gamma^h_{\rm NP}}\\
	&&\mu^{ggF}_{\gamma\gamma}\approx\frac{\Gamma_{\rm NP}(h\rightarrow gg)\Gamma_{\rm NP}(h\rightarrow \gamma\gamma)/\Gamma^h_{\rm NP}}{\Gamma_{\rm SM}(h\rightarrow gg)\Gamma_{\rm SM}(h\rightarrow \gamma\gamma)/\Gamma^h_{\rm SM}}=\frac{\Gamma_{\rm NP}(h\rightarrow gg)\Gamma_{\rm NP}(h\rightarrow \gamma\gamma)\Gamma^h_{\rm SM}}{\Gamma_{\rm SM}(h\rightarrow gg)\Gamma_{\rm SM}(h\rightarrow \gamma\gamma)\Gamma^h_{\rm NP}}
\end{eqnarray}
where $\Gamma^h_{\rm NP}$ and $\Gamma^h_{\rm SM}$ denote total decay widths in the NP model and the SM respectively.

\section{Numerical analysis}

In this section, we discuss the numerical results of the Higgs boson decays $h\rightarrow MZ$ in the TNMSSM are present. The results are constrained by the SM-like Higgs boson mass in the TNMSSM with $124.74\,{\rm GeV}\leq m_{{h}} \leq125.76\:{\rm GeV}$, where a $3\sigma$ experimental error is considered. For the SM parameters, we take $m_W=80.385\rm GeV$, $m_Z=91.1876\rm GeV$, $m_u=2.16\rm MeV$, $m_d=4.67\rm MeV$, $m_s=93.4\rm MeV$, $m_c=1.27\rm GeV$, $m_b=4.18\rm GeV$, $m_t=172.69\rm GeV$. For the squark sector, we take $m_{\tilde{Q}}=m_{\tilde{u}}=m_{\tilde{d}}=diag(M_Q,M_Q,M_Q)$ and $T_{u,d}=Y_{u,d}diag(A_Q,A_Q,A_t)$ for simplicity. According to the latest experimental data \cite{PDG}, We take $M_Q=2\rm TeV$, $A_Q=1.5\rm TeV$. And for the slepton sector, we take $m_{\tilde{l}}=m_{\tilde{e}}=2~\rm TeV$, $T_e=Y_ediag(A_e,A_e,A_e)$ and $A_e=1.5\rm TeV$. Then we take $\mu = 1~\rm TeV$, $ \tan\beta = 8$, $\tan\beta'=10$, $ \lambda =0.95$,$\kappa=0.9$, $ \chi_t = 0.4$, $T_{\Lambda_T} = 1.5~\rm TeV$, $T_\kappa=700~\rm GeV$, $T_\lambda=-700~\rm GeV$ and $\sqrt{v_T^2+\bar{v}_T^2}=2~\rm GeV$. We employ the following parameters as variable parameters in the numerical analysis
\begin{eqnarray}
	M_2,~\Lambda_T,~\chi_d,~A_t,~T_{\chi_t},~ T_{\chi_d}.\nonumber
\end{eqnarray}

And in our next numerical analysis we keep the lightest chargino always more than $800~\rm GeV$, all the mass of sleptons and squarks are more than $1900~\rm GeV$.
\subsection{The $h\rightarrow\gamma\gamma$, $h\rightarrow VV^*$ and$h\rightarrow Z\gamma$ in the TNMSSM}
\begin{figure}[h!]
\setlength{\unitlength}{1mm}
\centering
\subfigure{
\includegraphics[width=3in]{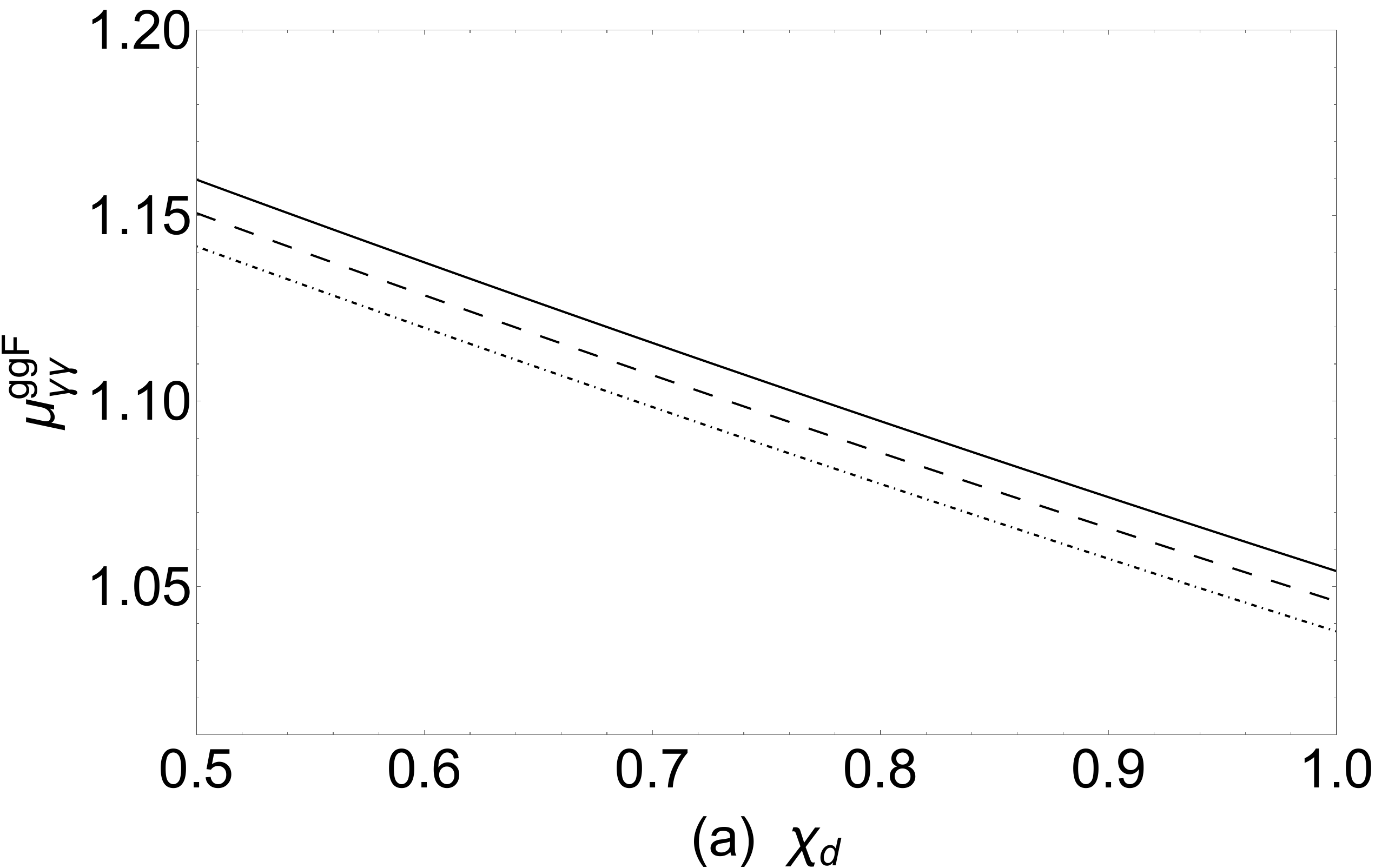}
\label{hto2photon}}
\subfigure{
\includegraphics[width=3in]{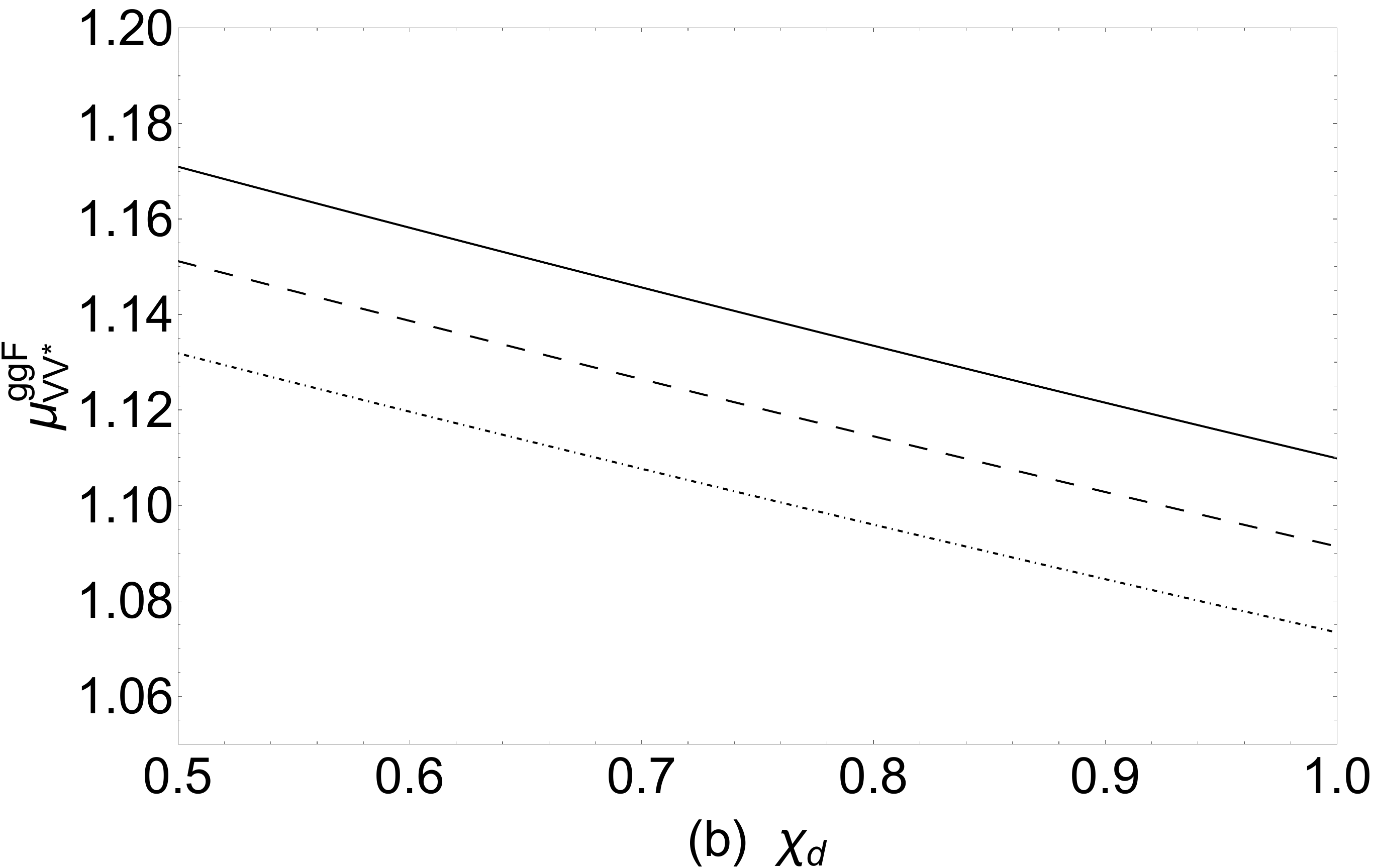}
\label{htovv}}
\subfigure{
\includegraphics[width=3in]{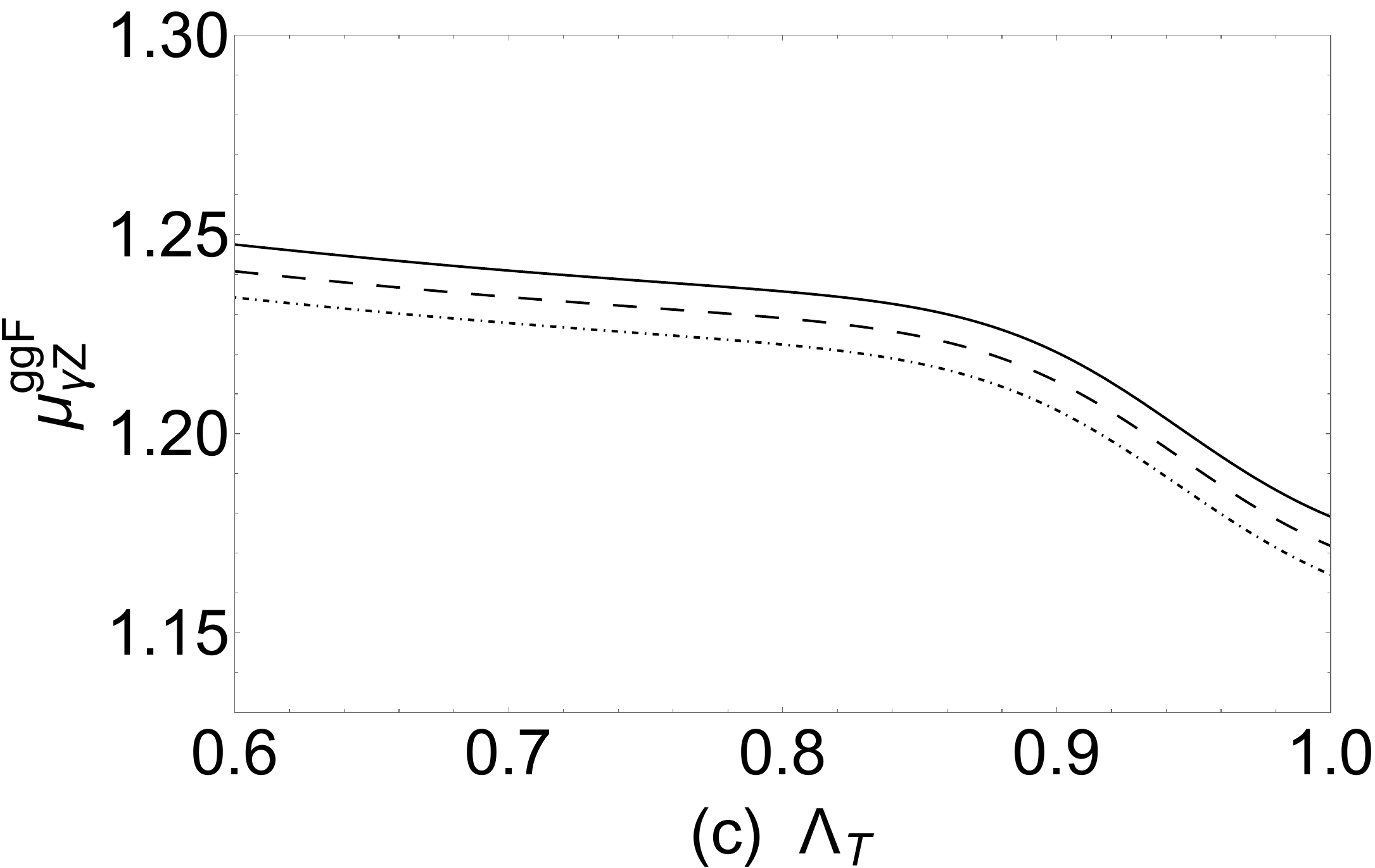}
\label{htoZgamma}}
\caption{(a) $\mu_{\gamma\gamma}^{ggF}$ varing with $\chi_d$ for $T_{\chi_t}=-800~\rm GeV$ (solid line), $T_{\chi_t}=-900~\rm GeV$ (dashed line) and $T_{\chi_t}=-1000~\rm GeV$ (dot dashed line). (b) $\mu_{VV^*}^{ggF}$ varying with $\chi_d$ for $\Lambda_T=0.7$ (solid line), $\Lambda_T=0.8$ (dashed line) and $\Lambda_T=0.9$ (dot dashed line). (c)  $\mu_{Z\gamma}^{ggF}$ varying with $\Lambda_T$ for $\chi_d=0.7$ (solid line), $\chi_d=0.8$ (dashed line) and $\chi_d=0.9$ (dot dashed line).}
\label{mu2gamma k}
\end{figure}

In this subsection we calculate the signal strengths for process $h\rightarrow\gamma\gamma$, $h\rightarrow VV^*$ and $h\rightarrow Z\gamma$. Some relevant formulas of $h\rightarrow\gamma\gamma$ and $h\rightarrow VV^*$ can be found in the works\cite{higgs decay1,higgs decay2}. At first we take parameters $M_2=1500~\rm Gev$, $\Lambda_T=0.8$, $A_t=1500~\rm GeV$ and $T_{\chi_d}=-800~\rm GeV$. And we paint the signal strength of the $h\rightarrow\gamma\gamma$ varying with $\chi_d$ in Fig.\ref{hto2photon}, for $T_{\chi_t}=-800~\rm GeV$ (solid line), $T_{\chi_t}=-900~\rm GeV$ (dashed line) and $T_{\chi_t}=-1000~\rm GeV$ (dot dashed line). In order to keep the SM-like Higgs mass satisfy the $3\sigma$ error of experimental constraints, we let the $\chi_d$ vary from $0.5$ to $1$. In Fig.\ref{hto2photon}, all the three curves are $1.03<\mu^{ggF}_{\gamma\gamma}<1.16$ and they they behave the same way. These curves tend to be decreases with the increase of the $\chi_d$. The solid line varies from 1.16 to 1.05, the dashed line varies from 1.15 to 1.045 and the dot dashed line varies from 1.14 to 1.035. Our results for process $h\rightarrow\gamma\gamma$ satisfy the experiment constraints\cite{PDG}.

Then, the signal strengths for processes $h\rightarrow ZZ^*$ and $h\rightarrow WW^*$ are very close. So we take $\mu^{ggF}_{VV^*}=\mu^{ggF}_{WW^*}=\mu^{ggF}_{ZZ^*}$ for simplicity, and we just paint the signal strength of $h\rightarrow ZZ^*$. We take parameters $M_2=1500~\rm GeV$, $T_{\chi_d}=-800~\rm GeV$, $T_{\chi_t}=-800~\rm GeV$ and $A_t=1500~\rm GeV$. And we paint the signal strength of the $h\rightarrow VV^*$ varying with $\chi_d$ in Fig.\ref{htovv}, for  $\Lambda_T=0.7$ (solid line), $\Lambda_T=0.8$ (dashed line) and  $\Lambda_T=0.9$ (dot dashed line). In Fig.\ref{htovv} the signal strength of the $h\rightarrow VV^*$ decreases with the increase of $\chi_d$. These curves are above 1.073 and below 1.171, and their behaviors are similar to each other. The experiment constraints\cite{PDG} $\mu^{(exp)}_{ZZ^*}=1.01\pm 0.07$, $\mu^{(exp)}_{WW^*}=1.19\pm 0.12$. So our calculate result $\mu^{ggF}_{ZZ^*}$ satisfies the experimental constraint that the error is $1\sigma$, and $\mu^{ggF}_{WW^*}$ satisfies the experimental constraint that the error is $2\sigma$.

The new physics contributions to the decay $h\rightarrow MZ$ come from the effective coupling of $hZ\gamma$. So we study the process $h\rightarrow Z\gamma$ at this subsection. In the numerical calculation of process $h\rightarrow Z\gamma$, we take the parameters $M_2=1500~\rm GeV$, $T_{\chi_d}=-800~\rm GeV$, $T_{\chi_t}=-800~\rm GeV$ and $A_t=1500~\rm GeV$. In Fig \ref{htoZgamma}, these curves are close to each other. And all the curves are varies from 1.164 to 1.248. When $0.6\leq \Lambda_T\leq 0.9$, all the lines here have a smaller slope. When $0.9\leq \Lambda_T\leq 1$, all the lines here have a bigger slope. And the result agrees with the observed signal strength with 1.5$\sigma$.

\subsection{The processes $h\rightarrow MZ$}
In this subsection we will study the processes $h\rightarrow MZ$. The vector mesons decay constants for $\omega$, $\rho$, $\phi$, $J/\psi$ and $\Upsilon$ can be found in Table \ref{t1}. The NP contribution of the process $h\rightarrow MZ$ comes from the effective coupling $hZ\gamma$. So our calculated results of decay $h\rightarrow MZ$ for different mesons should be similar.

\begin{figure}[h!]
\setlength{\unitlength}{1mm}
\centering
\subfigure{
	\includegraphics[width=2.5in]{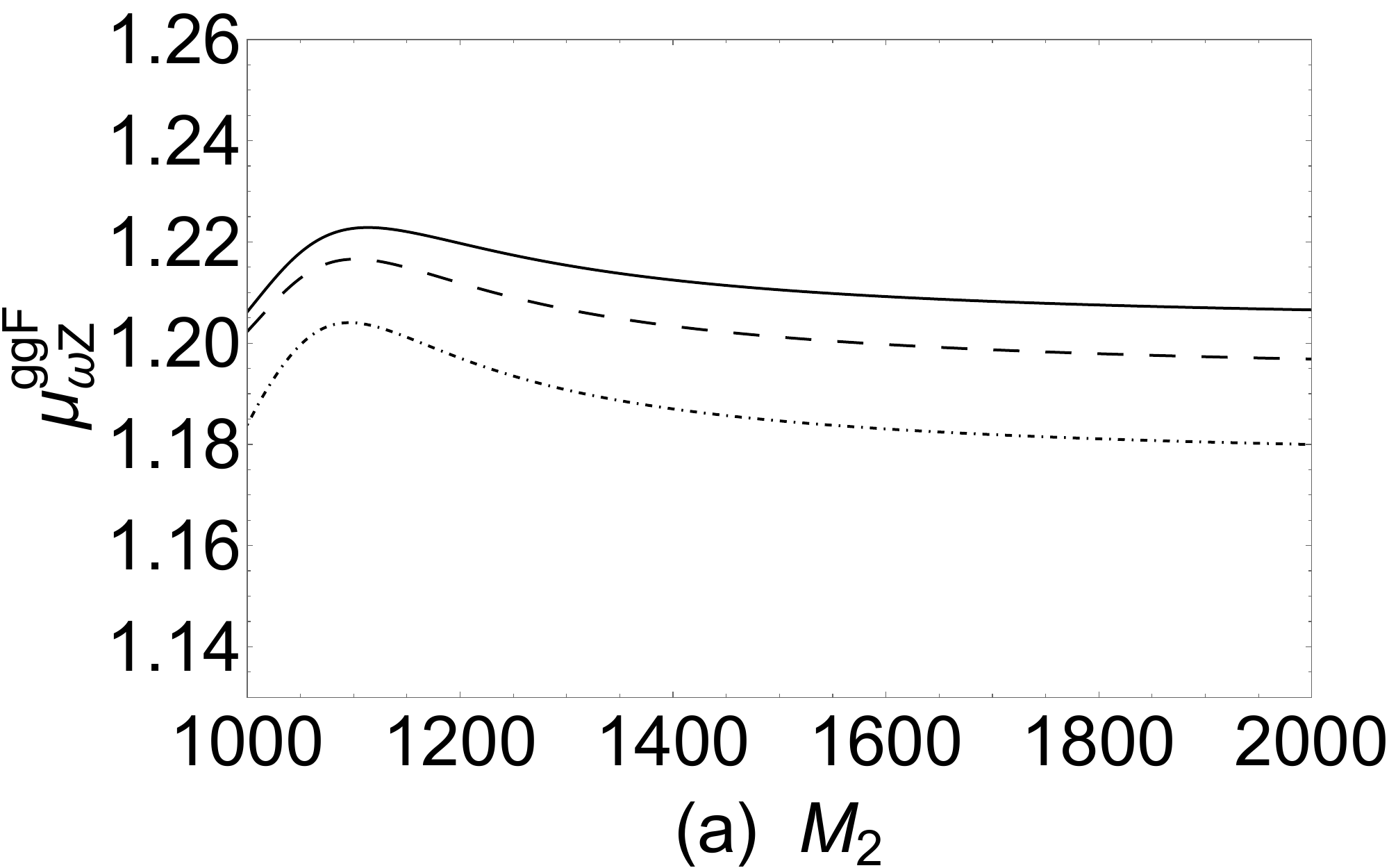}
	\label{hm1tcd}}
\subfigure{
	\includegraphics[width=2.5in]{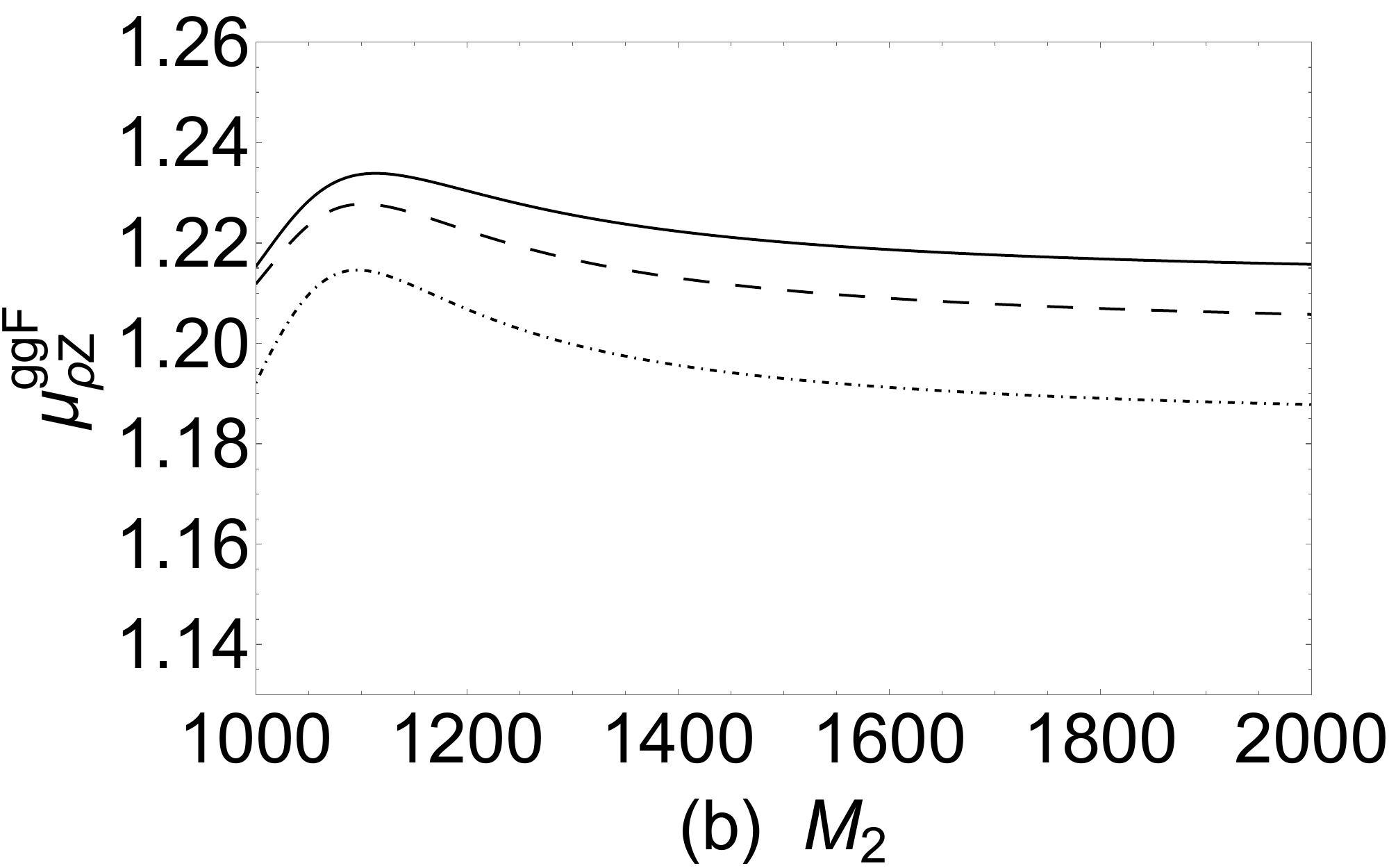}
	\label{hm2tcd}}
\subfigure{
	\includegraphics[width=2.5in]{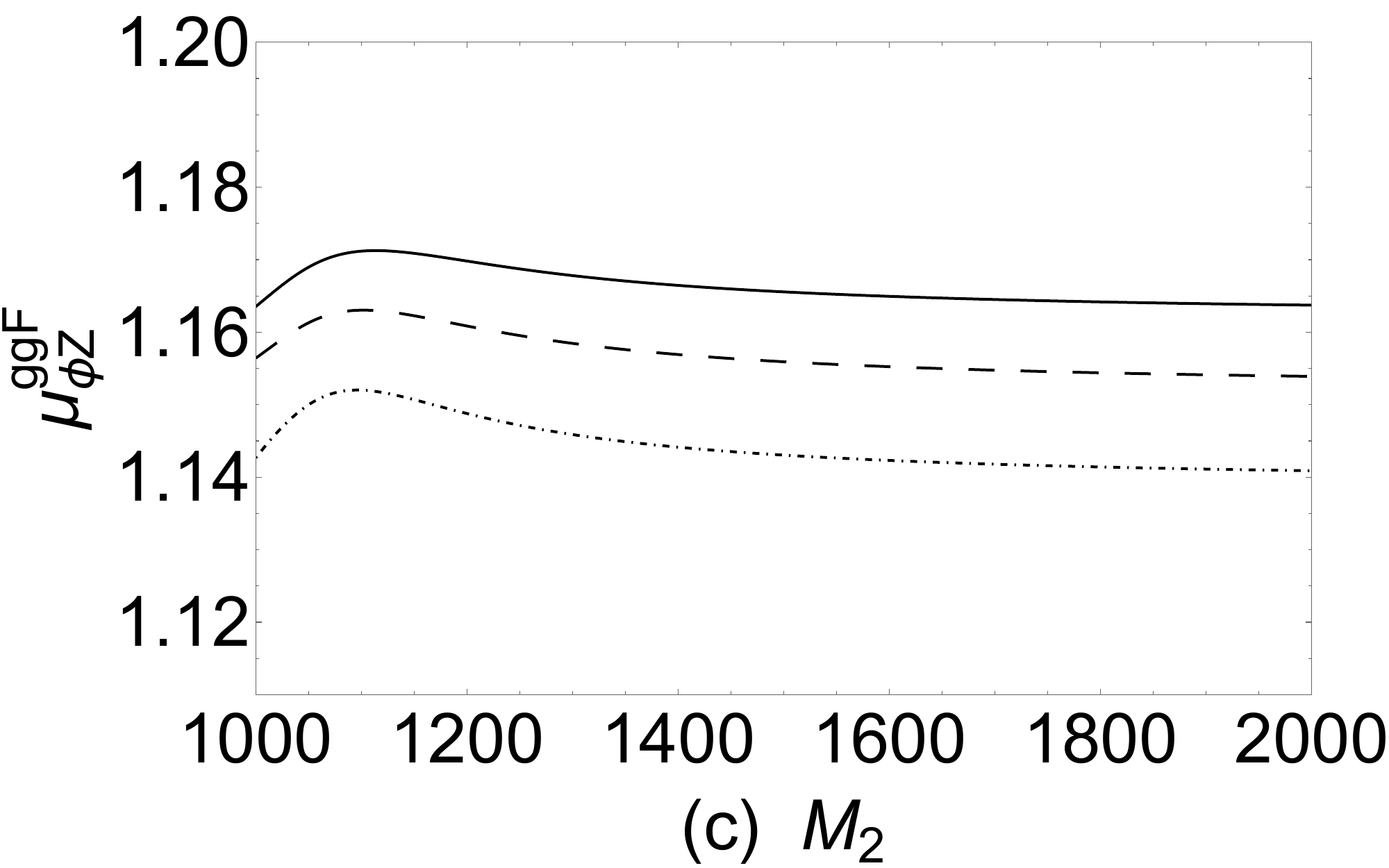}
	\label{hm3tcd}}
\subfigure{
	\includegraphics[width=2.5in]{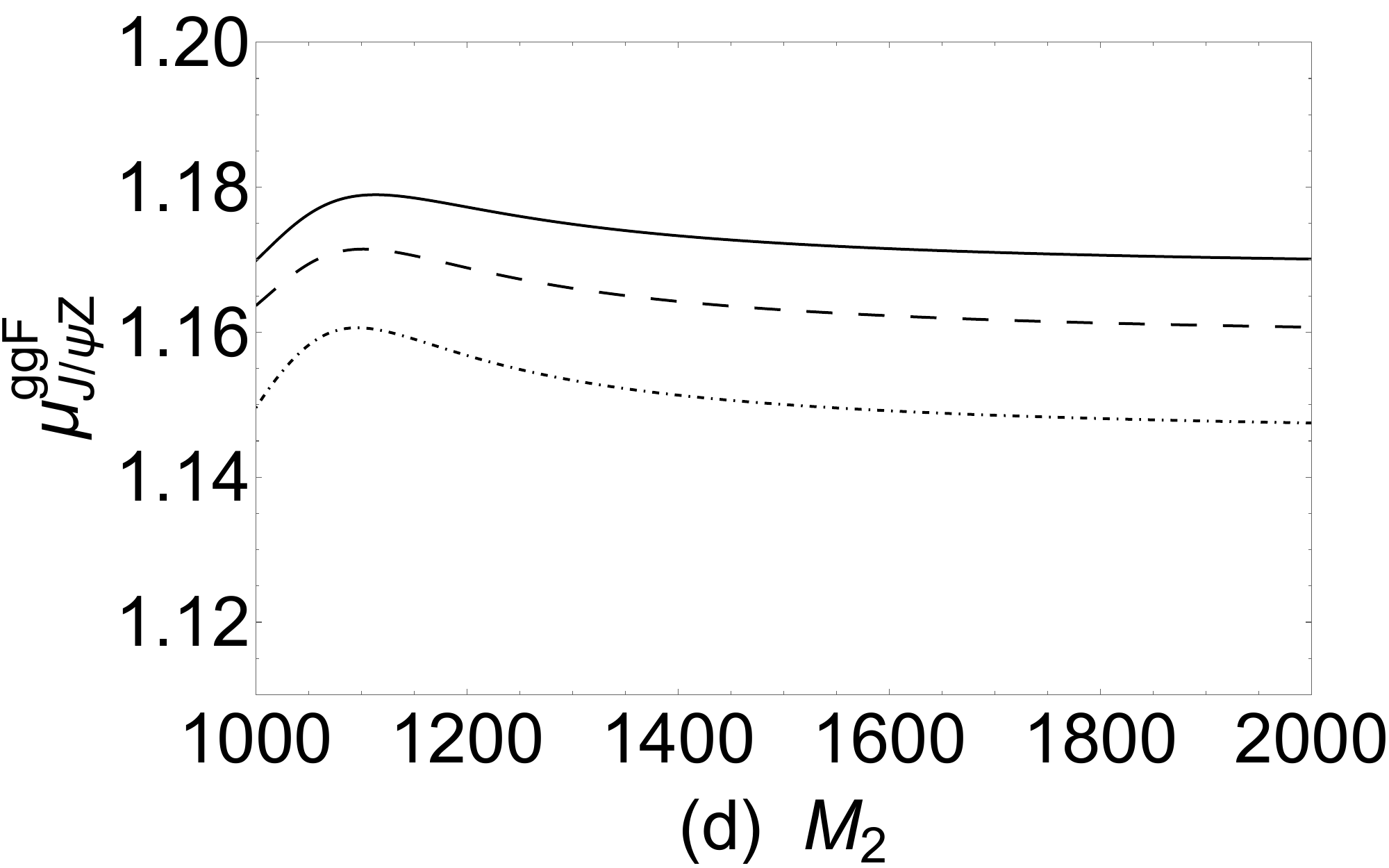}
	\label{hm4tcd}}
\subfigure{
	\includegraphics[width=2.5in]{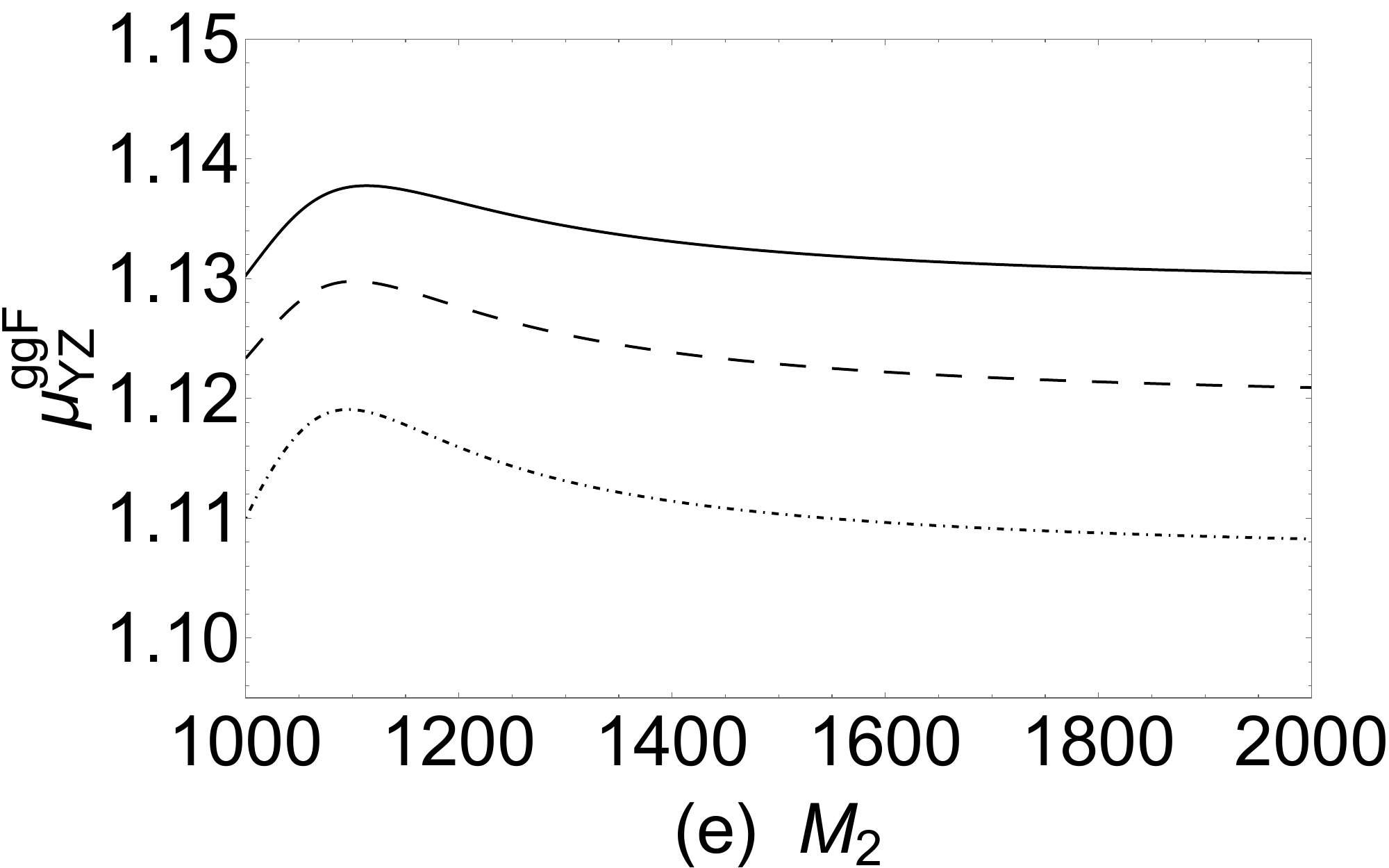}
	\label{hm5tcd}}
\caption[]{The signal strengths versus $M_2$ are plotted by the the solid line ($\chi_d=0.7$, $\Lambda_T=0.9$), dashed line ($\chi_d=0.8$, $\Lambda_T=0.8$) and the dot dashed line ($\chi_d=0.9$, $\Lambda_T=0.7$), respectively.}
\label{mu tcd}
\end{figure}
Now we study the signal strengths of process $h\rightarrow MZ$. First, we take the parameter $T_{\chi_d}=-800~\rm GeV$, $T_{\chi_t}=-800~\rm GeV$ and $A_t=1500~\rm GeV$. We paint the signal strengths of processes $h\rightarrow MZ$ in Fig.\ref{mu tcd}. And in Fig.\ref{mu tcd} the solid line is obtained with $\chi_d=0.7$, $\Lambda_T=0.9$ the dashed line is obtained with $\chi_d=0.8$, $\Lambda_T=0.8$, and the dot dashed line is obtained with $\chi_d=0.9$, $\Lambda_T=0.7$. We can see from Fig.\ref{mu tcd} the signal strengths increase with the increase of $M_2$ at $1000~{\rm GeV} \leq M_2\leq 1100~\rm GeV$, the signal strengths decrease with the increase of $M_2$ at $1100~{\rm GeV}\leq M_2\leq 2000~\rm GeV$. The signal strengths of processes $h\rightarrow\omega Z$ are in region 1.18-1.223, the signal strengths of processes $h\rightarrow\rho Z$ are in region 1.183-1.235, the signal strengths of processes $h\rightarrow\phi Z$ are in region 1.141-1.172, the signal strengths of processes $h\rightarrow J\psi Z$ are in region 1.147-1.179, the signal strengths of processes $h\rightarrow\Upsilon Z$ are in region 1.108-1.138.

\begin{figure}[h!]
\setlength{\unitlength}{1mm}
\centering
\subfigure{
	\includegraphics[width=2.5in]{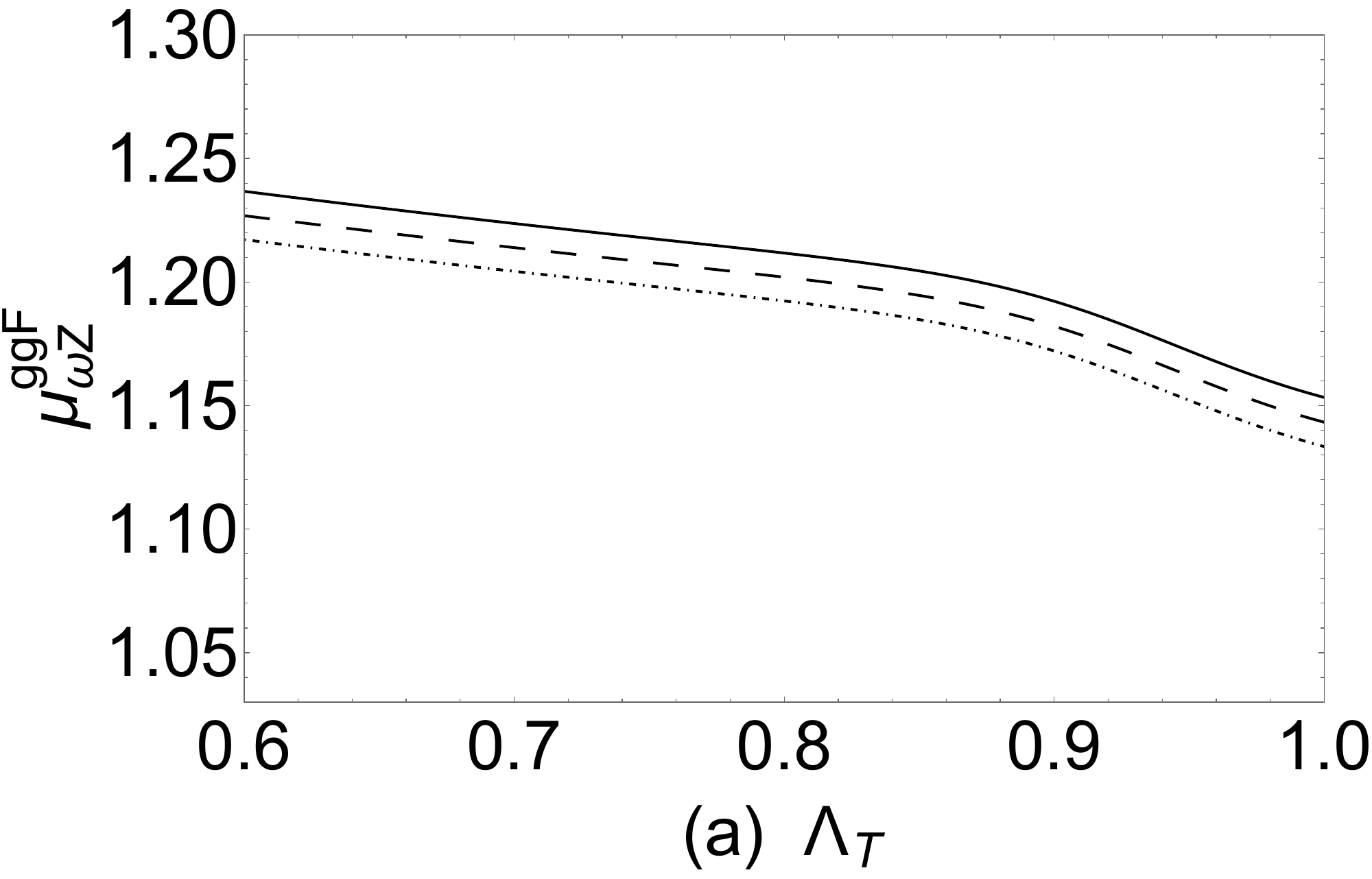}
	\label{hm1tl}}
\subfigure{
	\includegraphics[width=2.5in]{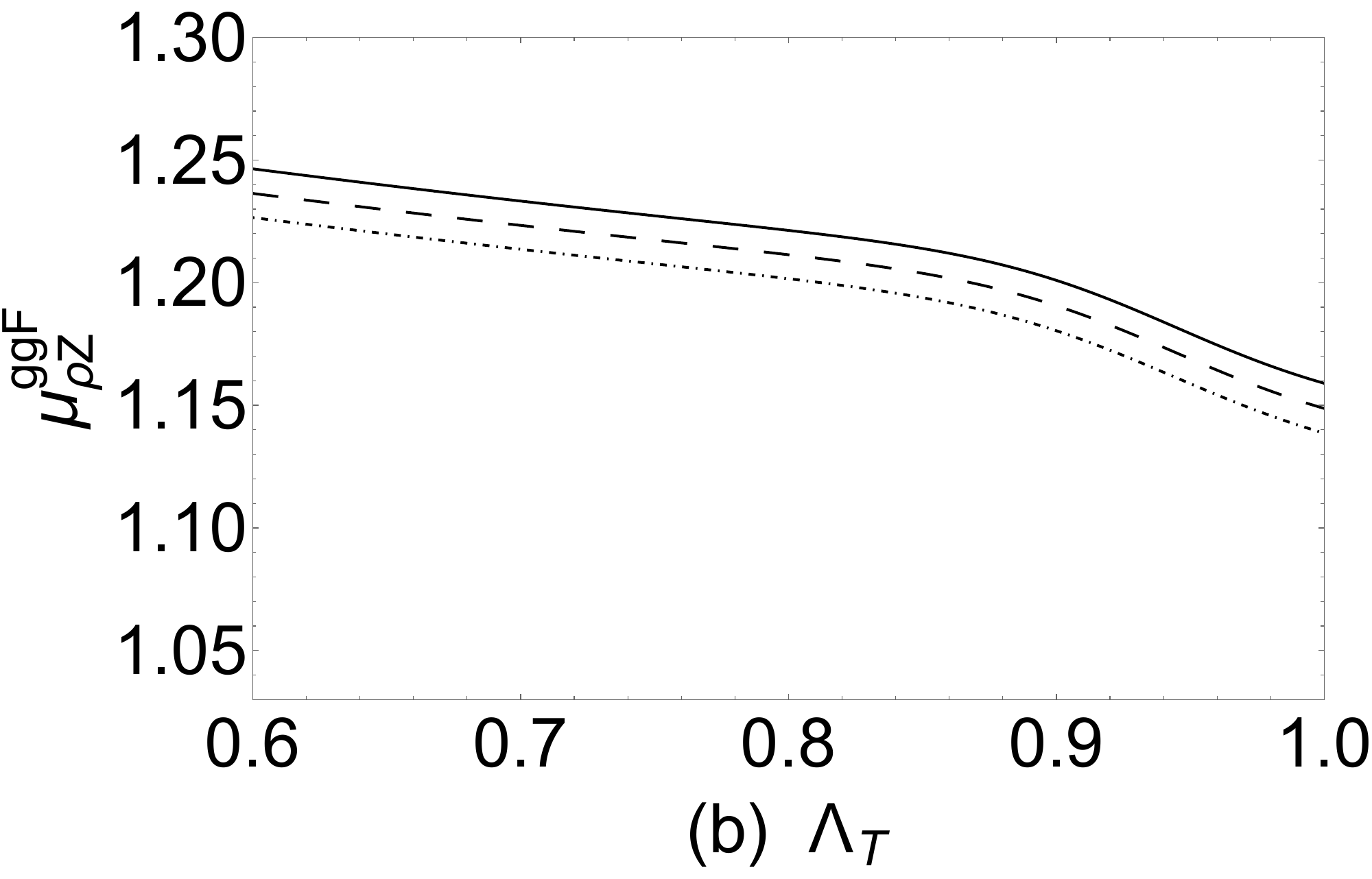}
	\label{hm2tl}}
\subfigure{
	\includegraphics[width=2.5in]{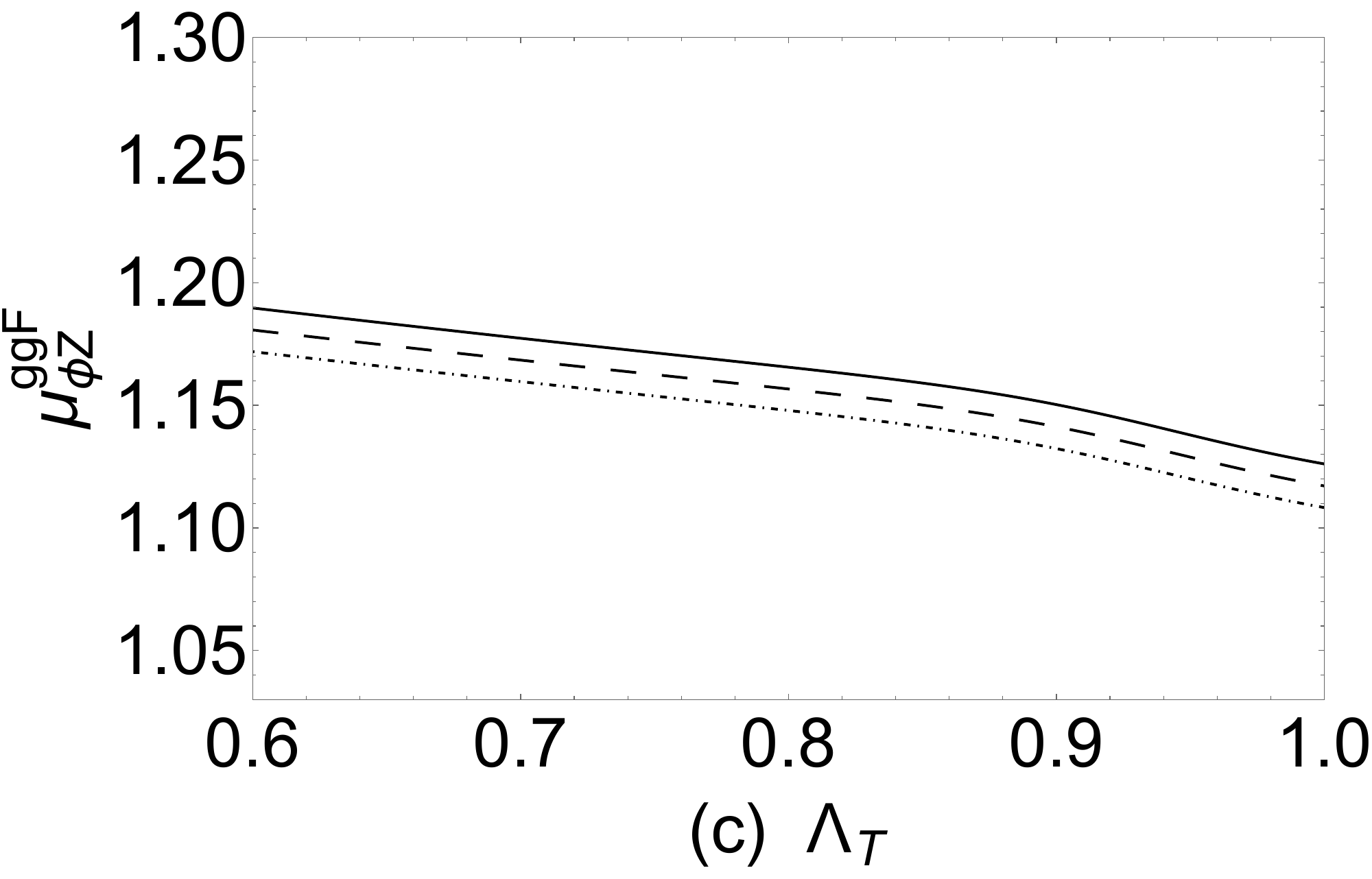}
	\label{hm3tl}}
\subfigure{
	\includegraphics[width=2.5in]{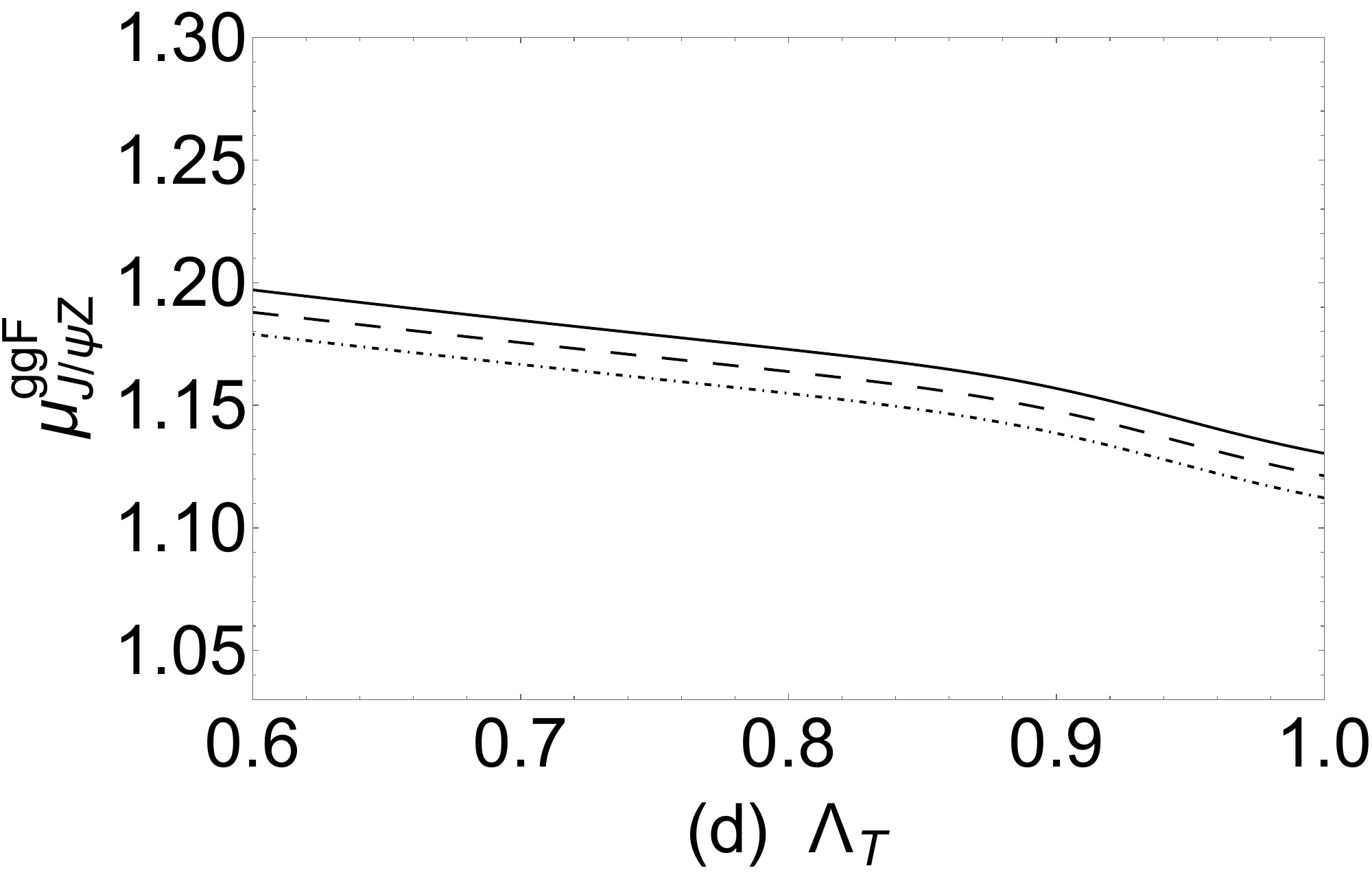}
	\label{hm4tl}}
\subfigure{
	\includegraphics[width=2.5in]{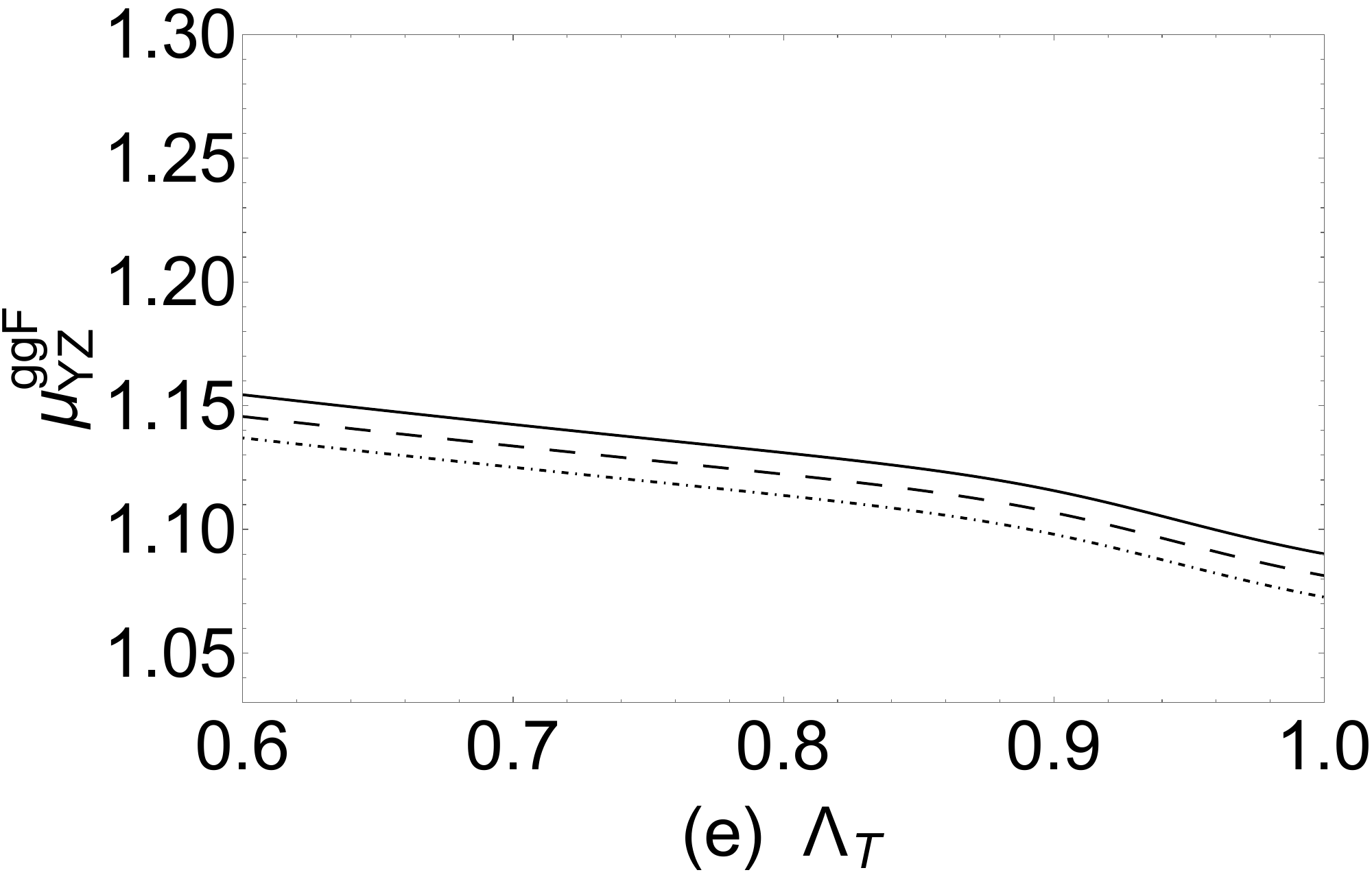}
	\label{hm5tl}}
\caption[]{The signal strengths versus $\Lambda_T$ are plotted by the solid line $(\chi_d=0.7)$, dashed line $(\chi_d=0.8)$ and dot dashed line $(\chi_d=0.9)$, respectively.}
\label{mu tl}
\end{figure}

The new contributions to the decay $h\rightarrow MZ$ come from the effective coupling of $h\gamma Z$. So we can infer that our results are consistent with the process $h\rightarrow\gamma Z$. As we can see, in Fig.\ref{htoZgamma}, the parameter $\Lambda_T$ has obvious influence on signal strength $\mu_{Z\gamma}^{ggF}$. So we should research the signal strengths of processes $h\rightarrow MZ$ versus $\Lambda_T$. The parameters we take as $M_2=1500~\rm GeV$, $T_{\chi_d}=-800~\rm GeV$, $T_{\chi_d}=-800~\rm GeV$ and $A_t=1500~\rm GeV$. And in order to keep the Higgs mass satisfy the $3\sigma$ error of experimental constraints, the $\Lambda_T$ is changed from $0.6$ to $1$ with $\chi_d=0.7,~0.8,~0.9$. The results for signal strengths of $\mu^{ggF}_{MZ}$ versus $\Lambda_T$ are plotted in Fig. \ref{mu tl}. The $\mu_{\omega Z}^{ggF}$ are in region 1.133-1.238, the $\mu_{\rho Z}^{ggF}$ are in region 1.138-1.249, the $\mu_{\phi_Z}^{ggF}$ are in region 1.109-1.191, the $\mu_{J/\psi Z}^{ggF}$ are in region 1.112-1.198 and the $\mu_{\Upsilon Z}^{ggF}$ are in region 1.07-1.155. And we can see that the signal strengths of $h\rightarrow MZ$ are similar to the signal strength of $h\rightarrow Z\gamma$. In Fig.\ref{mu tl}, these solid lines all the curves are varies from 1.164 to 1.248. When $0.6\leq \Lambda_T\leq 0.9$, all the lines here have a smaller slope. When $0.9\leq \Lambda_T\leq 1$, all the lines here have a bigger slope. And we can see the signal strengths of $h\rightarrow MZ$ are decrease as $\chi_d$ increase. 

\begin{figure}[h!]
\setlength{\unitlength}{1mm}
\centering
\subfigure{
	\includegraphics[width=2.5in]{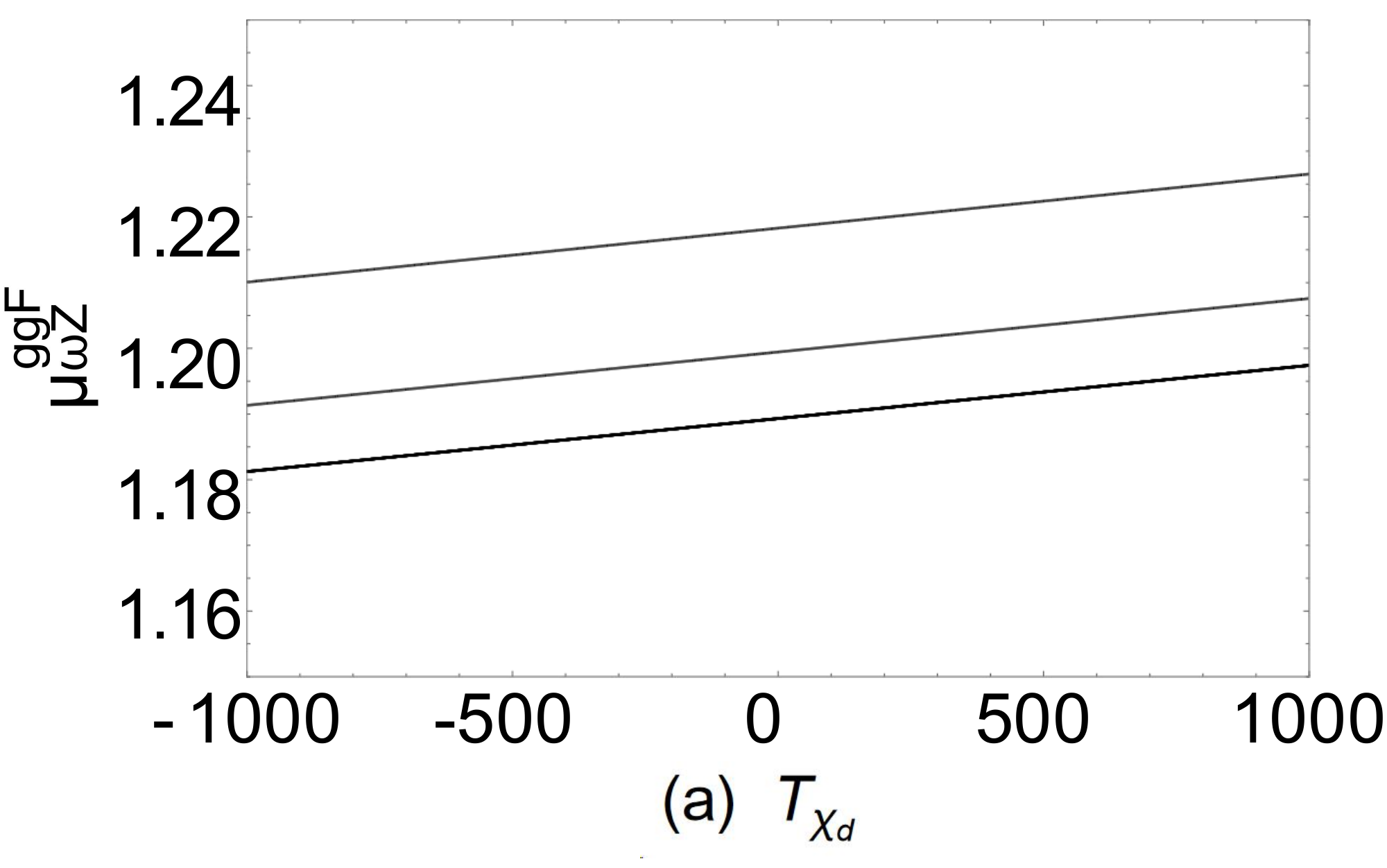}
	\label{m1 tk}}
\subfigure{
	\includegraphics[width=2.5in]{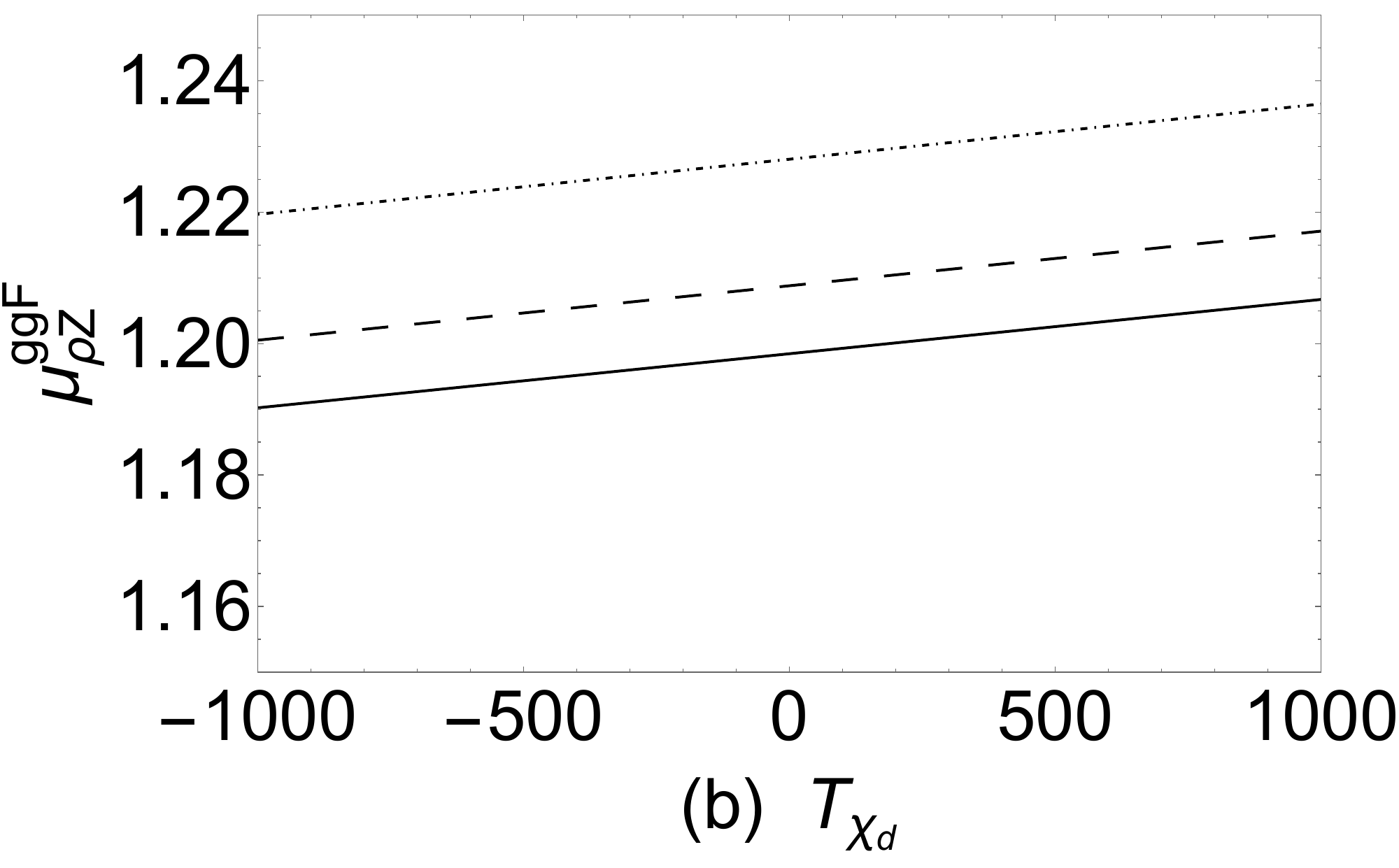}
	\label{m2 tk}}
\subfigure{
	\includegraphics[width=2.5in]{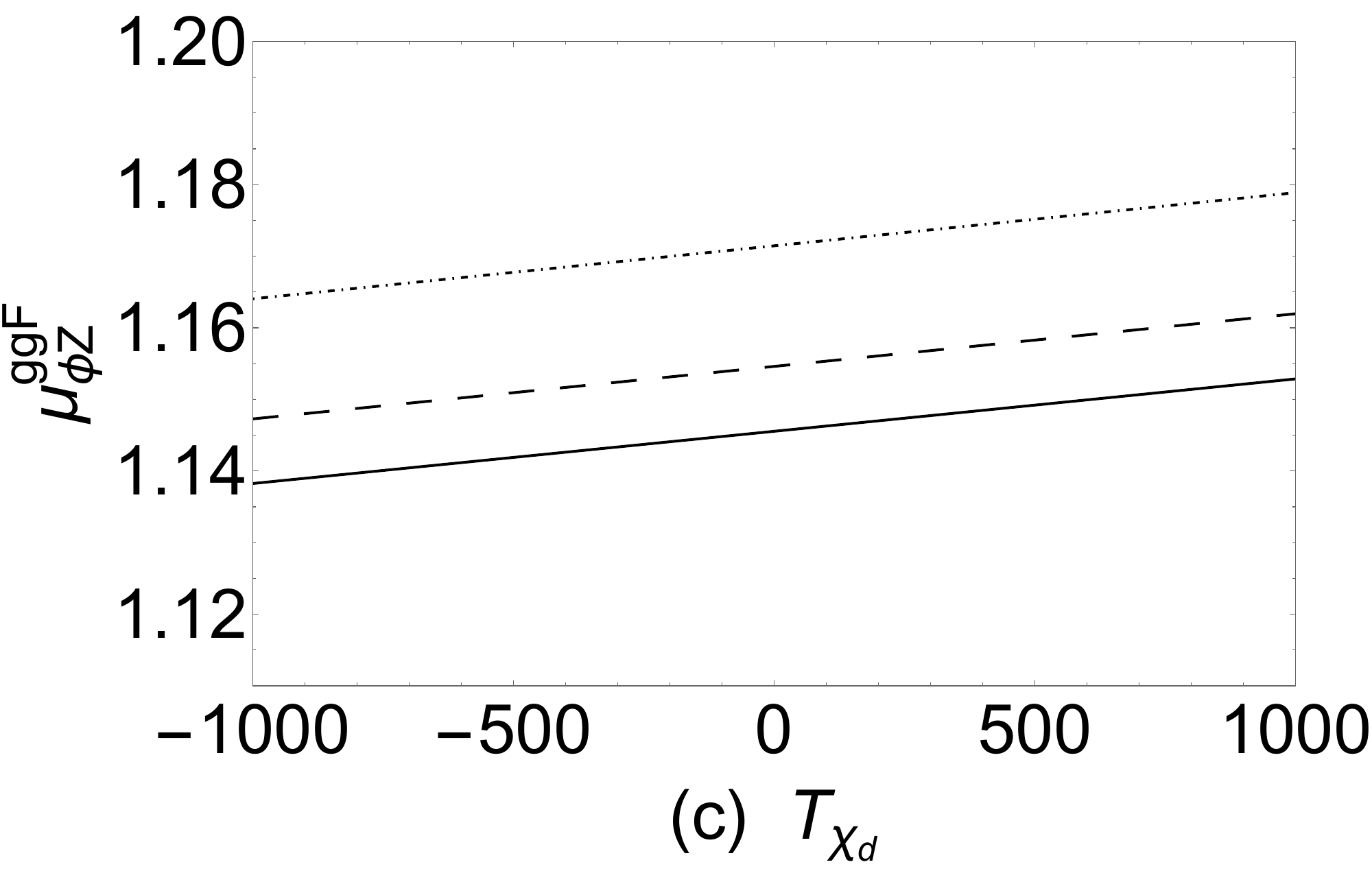}
	\label{m3 tk}}
\subfigure{
	\includegraphics[width=2.5in]{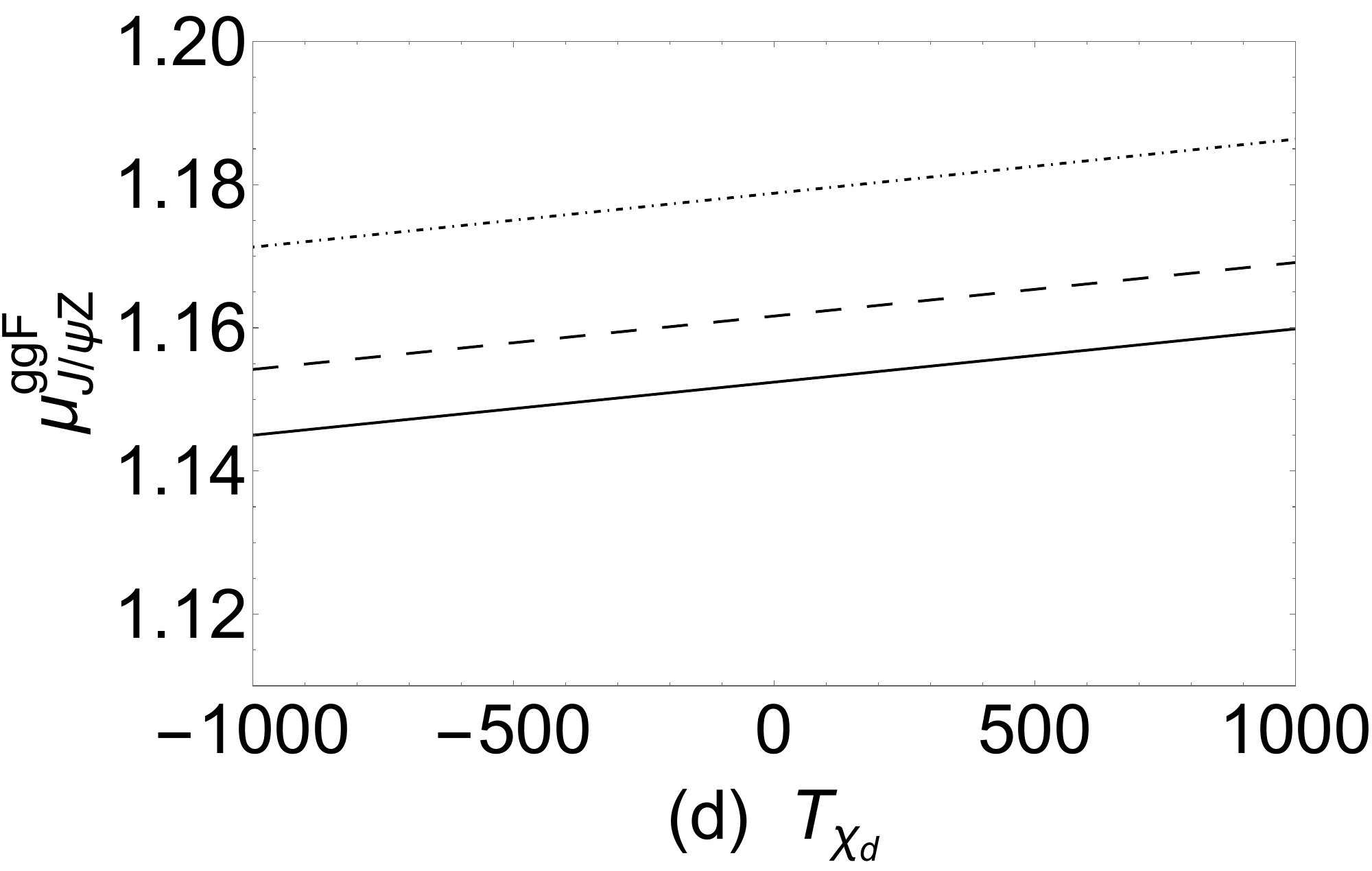}
	\label{m4 tk}}
\subfigure{
	\includegraphics[width=2.5in]{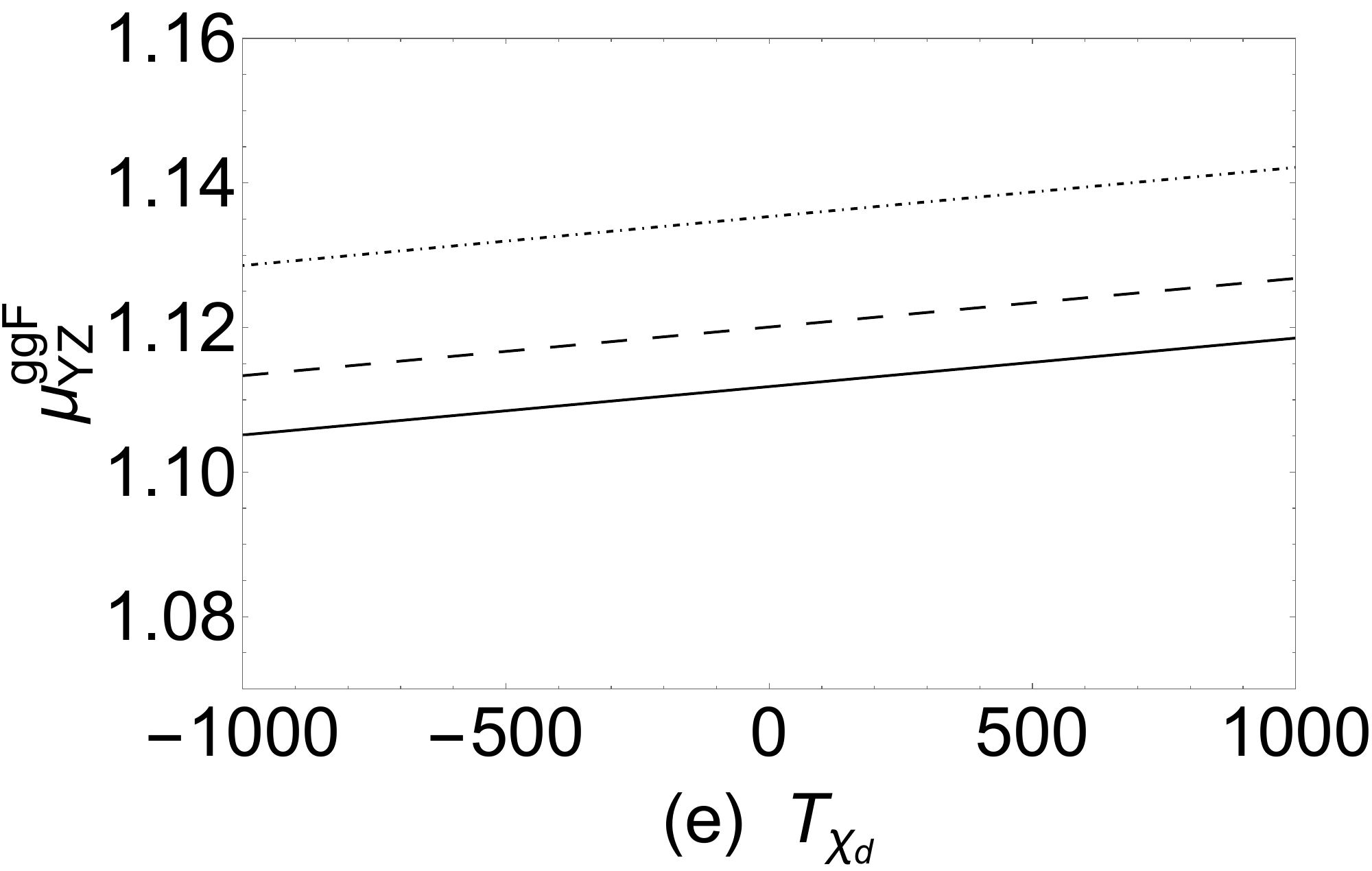}
	\label{m5 tk}}
	
\caption[]{The signal strengths versus $T_{\chi_d}$ are plotted by the solid line ($A_t=1000~\rm GeV$), dashed line ($A_t=1200~\rm GeV$) and dot dashed line ($A_t=1500~\rm GeV$).}
	
\label{mu tk}
\end{figure}

\begin{figure}[h!]
	\setlength{\unitlength}{1mm}
	\centering
	\subfigure{
		\includegraphics[width=2.5in]{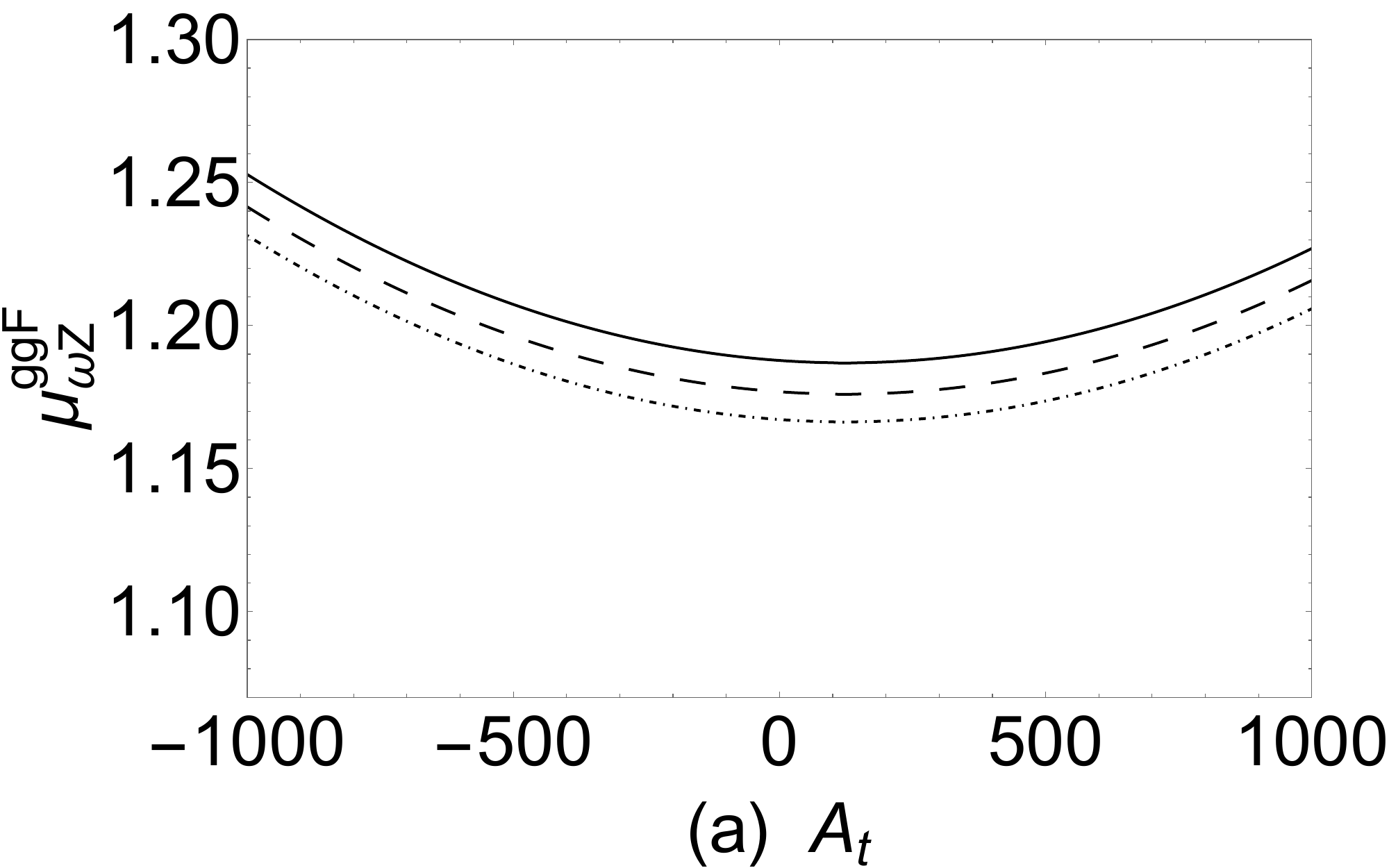}
		\label{m1 At}}
	\subfigure{
		\includegraphics[width=2.5in]{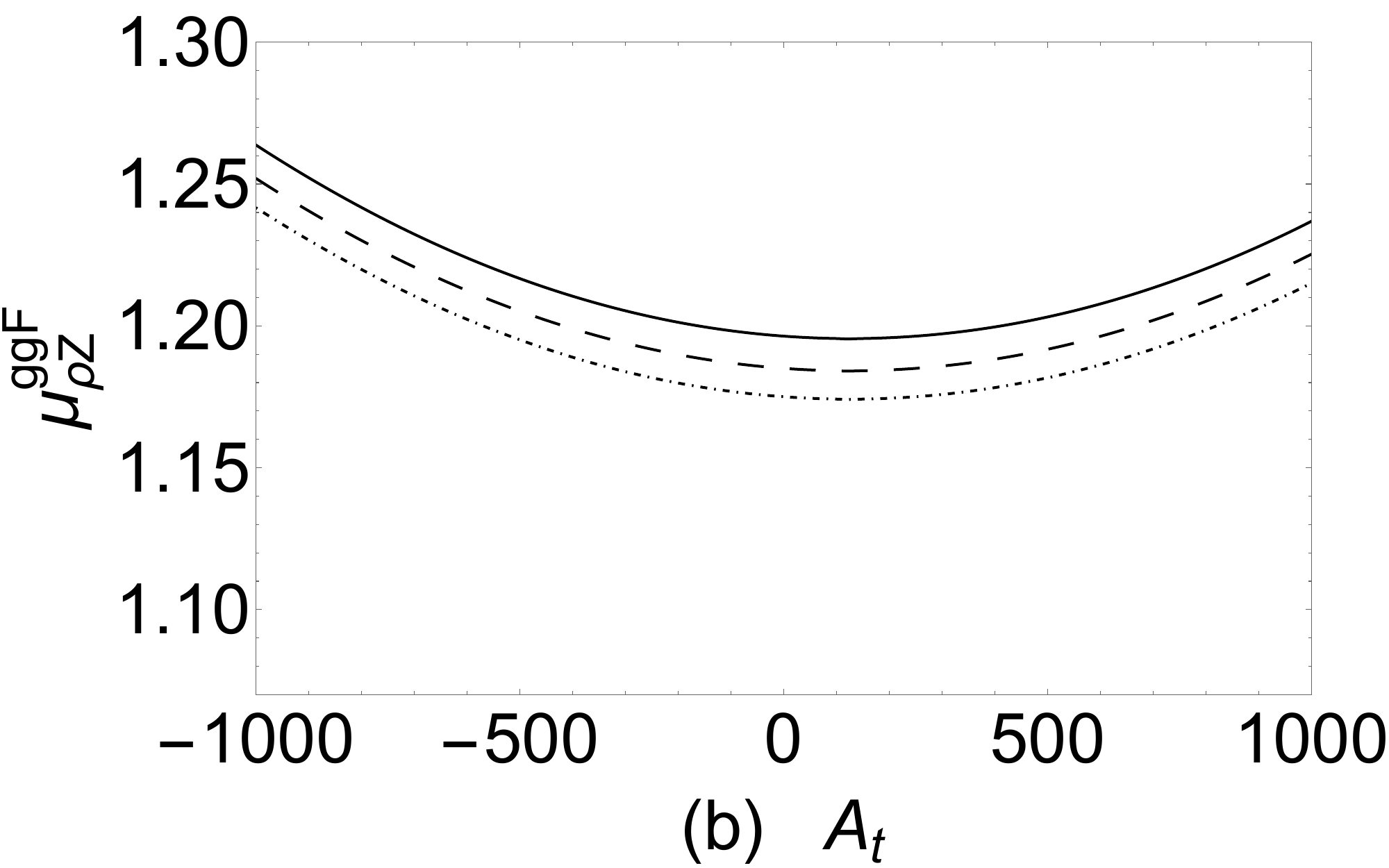}
		\label{m2 At}}
	\subfigure{
		\includegraphics[width=2.5in]{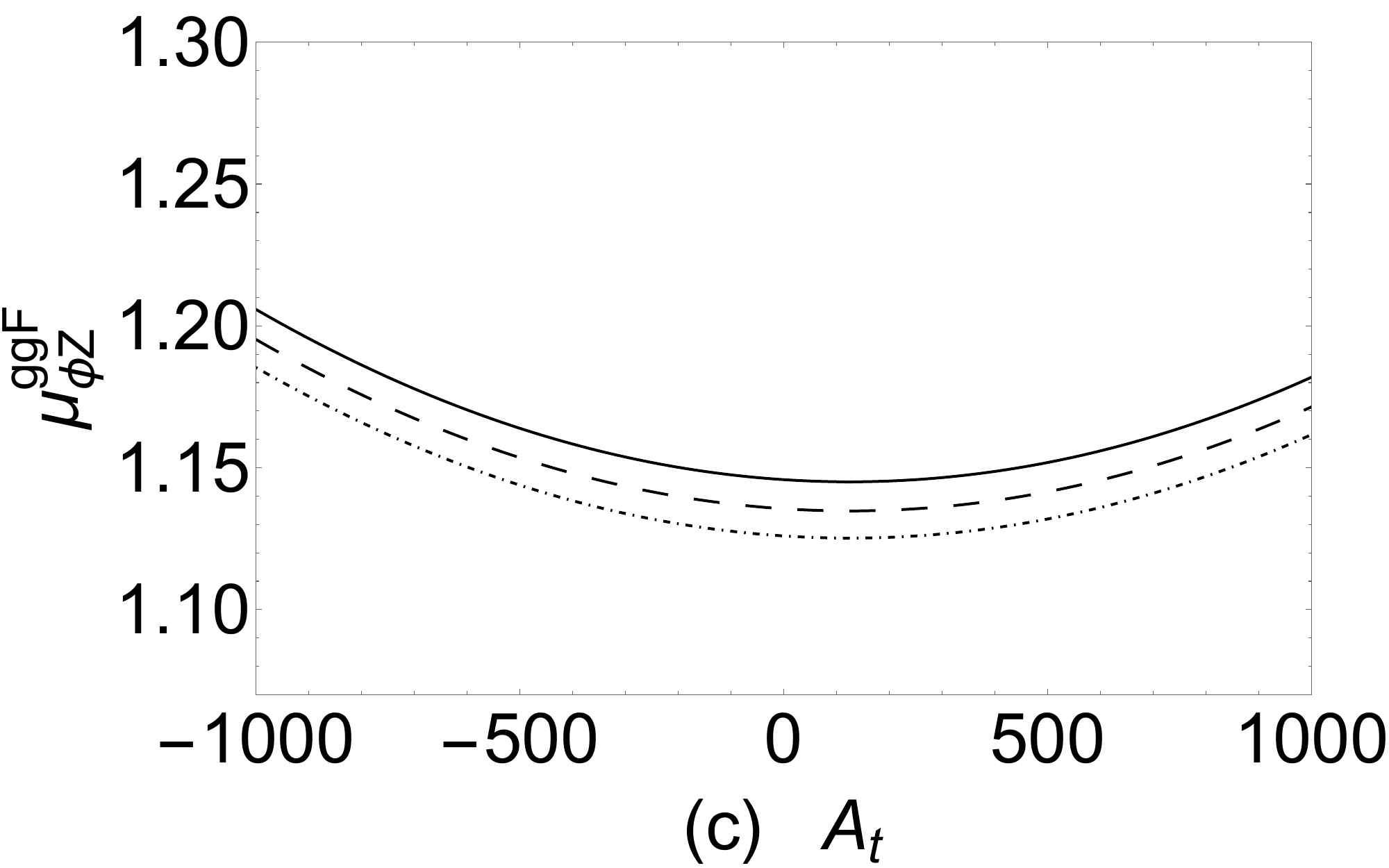}
		\label{m3 At}}
	\subfigure{
		\includegraphics[width=2.5in]{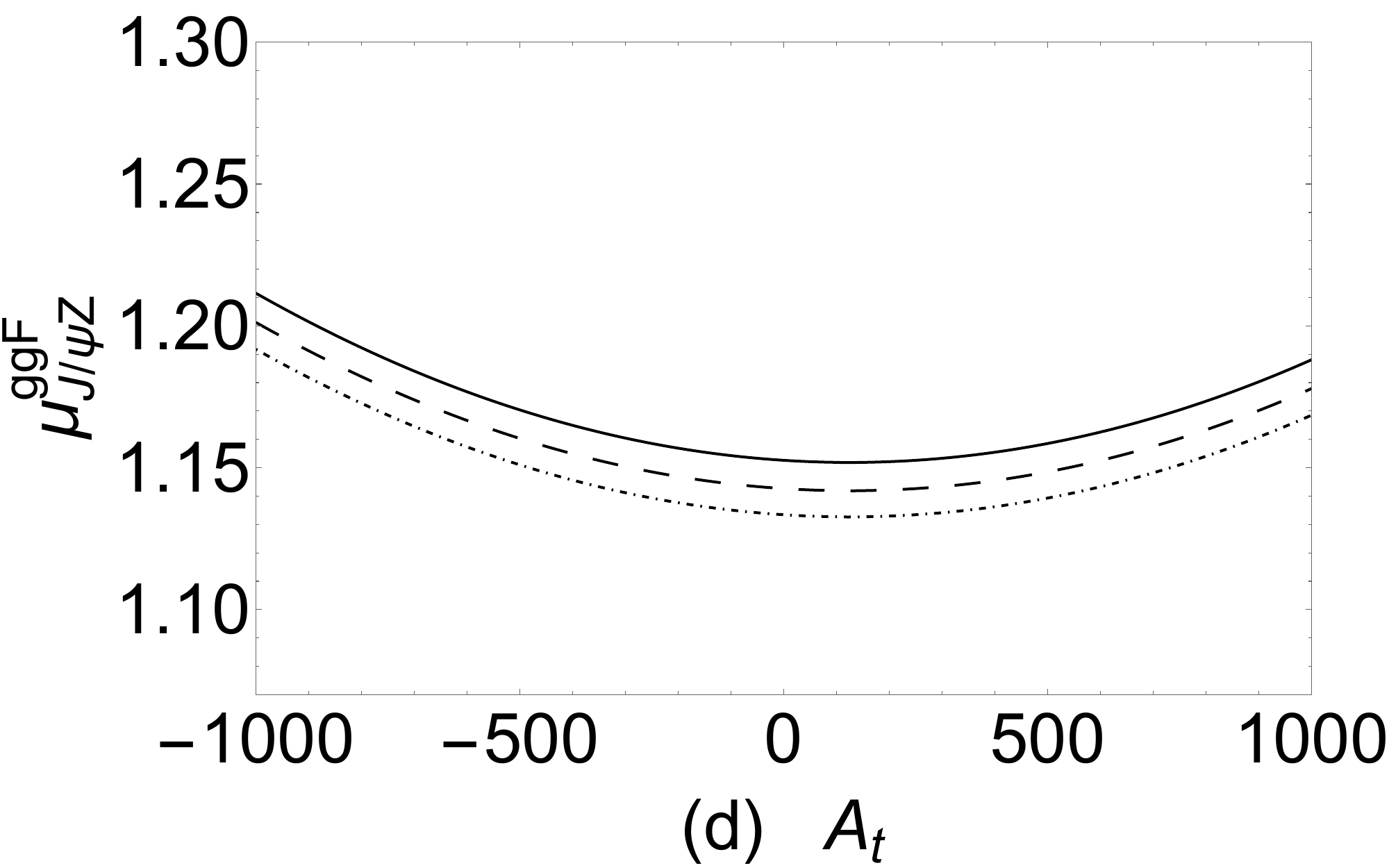}
		\label{m4 At}}
	\subfigure{
		\includegraphics[width=2.5in]{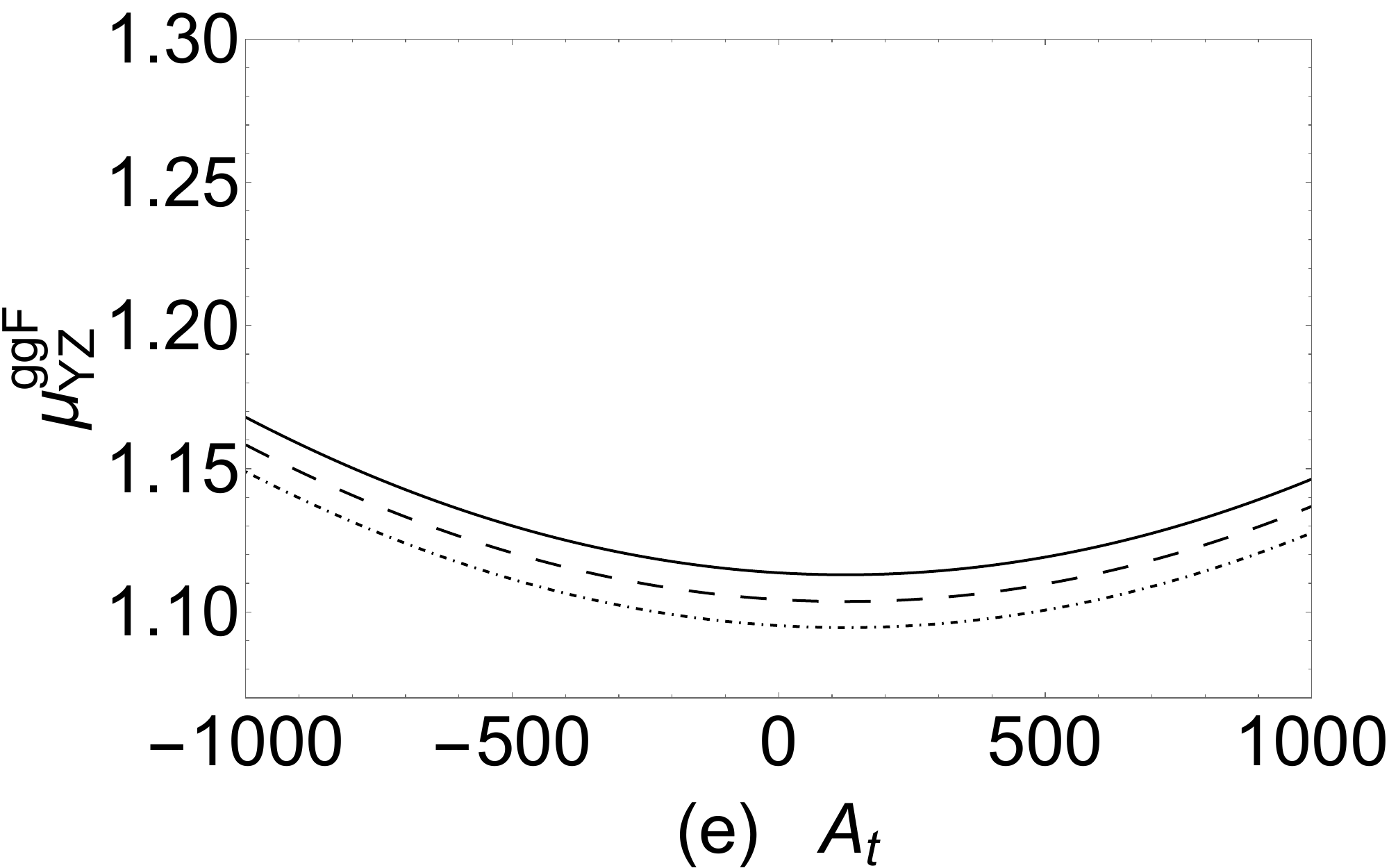}
		\label{m5 At}}
	\caption[]{The signal strengths versus $A_t$ are plotted by
			the solid line ($\Lambda_T=0.5$, $\chi_d=0.8$), dashed line ($\Lambda_T=0.6$, $\chi_d=0.7$) and dot dashed line ($\Lambda_T=0.7$, $\chi_d=0.6$), respectively.}
	\label{mu At}
\end{figure}

Then we study the effect of the $T_{\chi_d}$ on signal strengths of $h\rightarrow MZ$. The parameters we take as $M_2=1500~\rm GeV$, $\Lambda_T=0.8$, $\chi_d=0.7$ and $T_{\chi_t}=-800~\rm GeV$. And in order to keep the SM-like Higgs mass satisfy the $3\sigma$ error of experimental constraints, we let the $T_{\chi_d}$ vary from $-1000~\rm GeV$ to $1000~\rm GeV$ with $A_t=1000,~1200,~1500~\rm GeV$. We paint the signal strengths of processes $h\rightarrow MZ$ in Fig.\ref{mu tk}. And in Fig.\ref{mu tk} the solid lines are obtained with $A_t=1000~\rm GeV$, the dashed lines are obtained with $A_t=1200~\rm GeV$ and the dot dashed lines are obtained with $A_t=1500~\rm GeV$. In Fig.\ref{mu tk}, we can see tha the signal strengths increase with $T_{\chi_d}$ increase. In Fig.\ref{m1 tk} and Fig.\ref{m2 tk}, the signal strength of the processes of the $h\rightarrow\omega Z$ and $h\rightarrow\rho Z$ are in region 1.181-1.227 and 1.19-1.237. In Fig.\ref{m3 tk} and Fig.\ref{m4 tk}, the signal strength of processes $h\rightarrow\phi Z$ and $h\rightarrow J/\psi Z$ are region 1.138-1.178 and 1.145-1.186. The meson $\Upsilon$ is the heaviest meson we've studied. So that the signal strength of process $h\rightarrow\Upsilon Z$ is obviously less than the other results in Fig.\ref{mu tk}. The signal strength of process $h\rightarrow\Upsilon Z$ is in region 1.105-1.142. We can see from Fig.\ref{mu tk}, the $A_t$ have a great influence on the signal strengths of $h\rightarrow MZ$. The signal strengths of $h\rightarrow MZ$ increase as the $A_t$ increases.

At last we study the effect of the $A_t$ on signal strengths of processes $h\rightarrow MZ$. The coupling of higgs and third generation squarks include the $A_t$.  
We take the parameters as $M_2=1500~\rm GeV$,  $T_{\chi_d}=-800~\rm GeV$ and $T_{\chi_t}=-800~\rm GeV$. In order to keep the SM-like Higgs mass satisfy the $3\sigma$ error of experimental constrains, we let the $A_t$ vary from $-1000~\rm GeV$ to $1000~\rm GeV$ with $(\Lambda_T=0.5,~\chi_d=0.8)$, $(\Lambda_T=0.6,~\chi_d=0.7)$ and $(\Lambda_T=0.7,~\chi_d=0.6)$. We paint signal strengths of process $h\rightarrow MZ$ in Fig.\ref{mu At}. In Fig.\ref{mu At} the solid lines are obtained with $\Lambda_T=0.5$, $\chi_d=0.8$, the dashed lines are obtained with $\Lambda_T=0.6$, $\chi_d=0.7$ and the dot dashed lines are obtained with $\Lambda_T=0.7$, $\chi_d=0.6$. In Fig.\ref{m1 At} and Fig.\ref{m2 At}, the signal strengths of the processes $h\rightarrow\omega Z$ and $h\rightarrow\rho Z$ are in region 1.165-1.253 and 1.174-1.265. The signal strengths of processes $h\rightarrow\phi Z$ and $h\rightarrow J/\psi Z$ are in region 1.125-1.205 and 1.132-1.213. For the heaviest meson $\Upsilon$ we're studied, the signal strength of $h\rightarrow\Upsilon Z$ is obviously less than the other results in Fig.\ref{mu At}. The lines in Fig.\ref{m5 At} are in region 1.095-1.169.

\section{Conclusion}
In this work, we study the decays $h\rightarrow\gamma Z$ and $h\rightarrow MZ$ in the TNMSSM, with $M=\omega,\rho,\phi,J/\psi,\Upsilon$. There are two types of contributions to decay $h\rightarrow MZ$: the direct contributions and indirect contributions. For indirect contributions, there is a process $h\rightarrow Z\gamma^*\rightarrow MZ$, where $\gamma^*$ is off-shell and changes into the final state vector meson. There is no $h\gamma Z$ coupling at tree level, but it can be contributed by loop diagram. In the models beyond SM, the coupling constant can be divided into two parts: CP-even coupling constant $C_{\gamma Z}$ and CP-odd coupling constant $\tilde{C}_{\gamma Z}$. The CP-even coupling constant $C_{\gamma Z}$ is more important than the CP-odd coupling constant $\tilde{C}_{\gamma Z}$.

The experiment results of the signal strengths $\mu^{ggF}_{\gamma\gamma}$ and $\mu^{ggF}_{ZZ}$ are $\mu^{ggF}_{\gamma\gamma}=1.10 \pm 0.07$ and $\mu^{ggF}_{ZZ}=1.01 \pm 0.07$. Our numerical results of the signal strengths $\mu^{ggF}_{\gamma\gamma}$ and $\mu^{ggF}_{ZZ}$ are in region 1.035-1.16 and 1.073-1.171 which satisfy the error of $1\sigma$. Our numerical results of the signal strength in region 1.164-1.248. The result agrees with the observed signal strength with 1.5$\sigma$. The numerical results show that the TNMSSM contributions to the processes $h\rightarrow\omega Z$ and $h\rightarrow\rho Z$ are more considerable. The signal strengths $\mu^{ggF}_{\omega Z,\rho Z}$ are about 1.13-1.26. The TNMSSM corrections to the processes $h\rightarrow\phi Z$ and $h\rightarrow J/\psi Z$ are during 1.11-1.21, and $h\rightarrow\Upsilon Z$ about 1.07-1.17. The decays $h\rightarrow MZ$ may be accessible at future high energy colliders.

\begin{acknowledgments}
	The work has been supported by the National Natural Science Foundation of China (NNSFC) with Grants No. 12075074, No. 12235008, Natural Science Foundation of Guangxi Autonomous Region with Grant No. 2022GXNSFDA035068, Hebei Natural Science Foundation with Grant No. A2022201017, No. A2023201041, the youth top-notch talent support program of the Hebei Province.	
\end{acknowledgments}

\appendix
\section{Form factors }
The form factors are
\begin{eqnarray}
	A_0(\tau,\lambda)&&=I_1(\tau,\lambda),\\
	A_{1/2}(\tau,\lambda)&&=I_1(\tau,\lambda)-I_2(\tau,\lambda)
\end{eqnarray}
with
\begin{eqnarray}
	&&I_1(\tau,\lambda)=\frac{\tau\lambda}{2(\tau-\lambda)}+\frac{\tau^2 \lambda^2}{2(\tau-\lambda)^2}[f(\tau^{-1})-f(\lambda^{-1})]+\frac{\tau^2\lambda}{(\tau-\lambda)^2}[g(\tau^{-1})-g(\lambda^{-1})]\\
	&&I_2(\tau,\lambda)=-\frac{\tau \lambda}{2(\tau-\lambda)}[f(\tau^{-1})-f(\lambda^{-1})]
\end{eqnarray}
the $f(x)$ and $g(x)$ are here:
\begin{eqnarray}
	&&f(x)=\left\{\begin{array}{ll}
		\arcsin ^{2} \sqrt{x}, & x \leq 1 \\
		-\frac{1}{4}\left[\ln \frac{1+\sqrt{1-1 / x}}{1-\sqrt{1-1 / x}}-i \pi\right]^{2}, & x>1
	\end{array}\right.\\
	&&g(x)=\left\{\begin{array}{ll}
		\sqrt{x^{-1}-1}\arcsin \sqrt{x}, & x \ge 1 \\
		\frac{\sqrt{1-x^{-1}}}{2}\left[\ln \frac{1+\sqrt{1-1 / x}}{1-\sqrt{1-1 / x}}-i \pi\right], & x<1
	\end{array}\right.
\end{eqnarray}

\section{The mass of Higgs and Charginos}
In the basis $(\phi_d,\phi_u,\phi_s,\phi_T,\phi_{\bar{T}})$, the definition of mass squared matrix for neutral Higgs is given by
\begin{eqnarray}
	m_h^2=\begin{pmatrix}
		m_{\phi_d \phi_d}&m_{\phi_u \phi_d}&m_{\phi_s \phi_d}&m_{\phi_T \phi_d}&m_{\phi_{\bar{T}} \phi_d}\\
		m_{\phi_d \phi_u}&m_{\phi_u \phi_u}&m_{\phi_s \phi_u}&m_{\phi_T \phi_u}&m_{\phi_{\bar{T}} \phi_u}\\
		m_{\phi_d \phi_s}&m_{\phi_u \phi_s}&m_{\phi_s \phi_s}&m_{\phi_T \phi_s}&m_{\phi_{\bar{T}} \phi_s}\\
		m_{\phi_d \phi_T}&m_{\phi_u \phi_T}&m_{\phi_s \phi_T}&m_{\phi_T \phi_T}&m_{\phi_{\bar{T}} \phi_T}\\
		m_{\phi_d \phi_{\bar{T}}}&m_{\phi_u \phi_{\bar{T}}}&m_{\phi_s \phi_{\bar{T}}}&m_{\phi_T \phi_{\bar{T}}}&m_{\phi_{\bar{T}} \phi_{\bar{T}}}\\
	\end{pmatrix}
\end{eqnarray}
where
\begin{align}
	m_{\phi_d \phi_d}&=m_{H_d}^2+\frac{1}{8}(g_1^2+g_2^2)(2v_{\bar{T}}^2-2v_T^2+3v_d^2-v_u^2)\nonumber\\
	&+\sqrt{2}v_T\Re(T_{\chi_d})+\frac{|\lambda|^2}{2}(v_s^2+v_u^2)-v_s v_{\bar{T}} \Re(\chi_d \Lambda_T^*)+(3v_d^2+2v_T^2)|\chi_d|^2,\\	
	m_{\phi_d \phi_u}&=-\frac{1}{4}(g_1^2+g_2^2)v_d v_u-\frac{1}{\sqrt{2}}v_s \Re(T_\lambda)
	+\frac{1}{2}v_T v_{\bar{T}} \Re(\lambda\Lambda_T^*)+v_d v_u |\lambda|^2\nonumber\\
	&-v_s v_{\bar{T}}\Re(\chi_t \lambda^*)-v_s v_T\Re(\chi_d \lambda^*)-\frac{1}{2}v_s^2\Re(\kappa\lambda^*),\\
	m_{\phi_u \phi_u}&=m_{H_u}^2-\frac{1}{8}(g_1^2+g_2^2)(2v_{\bar{T}}^2-2v_T^2-3v_u^2+v_d^2)\nonumber\\
	&+\sqrt{2}v_{\bar{T}}\Re(T_{\chi_t})+\frac{1}{2}(v_d^2+v_s^2)|\lambda|^2-v_s v_T \Re(\chi_t \Lambda_T^*)+(2v_{\bar{T}}^2+3v_u^2)|\chi_t|^2,\\
	m_{\phi_d \phi_s}&=v_d v_s|\lambda|^2-\frac{1}{\sqrt{2}}v_u \Re(T_\lambda)-v_s v_u \Re(\kappa\lambda^*)
	-v_T v_u \Re(\lambda\chi_d^*)-v_{\bar{T}}v_u \Re(\chi_t \lambda^*)-v_d v_{\bar{T}} \Re(\Lambda_T \chi_d^*),\\
	m_{\phi_u \phi_s}&=v_s v_u|\lambda|^2-\frac{1}{\sqrt{2}}v_d\Re(T_\lambda)-v_d v_s\Re(\lambda\kappa^*)
	-v_T v_d\Re(\lambda\chi_d^*)-v_d v_{\bar{T}}\Re(\lambda\chi_t^*)-v_T v_u\Re(\chi_t \Lambda_T^*),\\
	m_{\phi_s \phi_s}&=m_S^2+\frac{|\Lambda_T|^2}{2}(v_T^2+v_{\bar{T}}^2)-v_T 
	v_{\bar{T}}\Re(\kappa\Lambda_T^*)+3v_s^2|\kappa|^2-v_d v_u\Re(\lambda\kappa^*)+\frac{1}{2}(v_d^2+v_u^2)|\lambda|^2,\nonumber\\
	&+\sqrt{2}v_s\Re(T_\kappa),\\
	m_{\phi_d \phi_T}&=-\frac{1}{2}(g_1^2+g_2^2)v_d v_u+\frac{1}{2}v_u v_{\bar{T}}\Re(\lambda\Lambda_T^*)-v_u 
	v_s\Re(\lambda\chi_d^*)+4v_d v_T|\chi_d|^2+\sqrt{2}v_d\Re(T_{\chi_d}),\\
	m_{\phi_u \phi_T}&=\frac{1}{2}(g_1^2+g_2^2)v_d v_T-v_u v_s\Re(\Lambda_T\chi_t^*)-v_d 
	v_s\Re(\lambda\chi_d^*)+\frac{1}{2}v_d v_{\bar{T}}\Re(\lambda\Lambda_T^*),
\end{align}
\begin{align}
	m_{\phi_s \phi_T}&=v_s v_T|\lambda|^2-\frac{1}{\sqrt{2}}v_{\bar{T}}\Re(T_\Lambda)-
	v_s v_{\bar{T}}\Re(\kappa\Lambda_T^*)-v_d v_u \Re(\lambda\chi_d^*)-\frac{1}{2}v_u^2\Re(\chi_t\Lambda_T^*),\\
	m_{\phi_T,\phi_T}&=m_T^2-\frac{1}{4}(g_1^2+g_2^2)(2v_{\bar{T}}^2-6v_T^2-v_u^2+v_d^2)+2v_d^2|\chi_d|^2
	+\frac{1}{2}(v_s^2+v_{\bar{T}}^2)|\Lambda_T|^2,\\
	m_{\phi_d,\phi_{\bar{T}}}&=\frac{1}{2}(g_1^2+g_2^2)v_d v_{\bar{T}}+\frac{1}{2}v_T v_u \Re(\lambda\Lambda_T^*)
	-v_d v_s \Re(\chi_d\Lambda_T^*)-v_s v_u\Re(\chi_t\lambda^*),\\
	m_{\phi_u,\phi_{\bar{T}}}&=-\frac{1}{2}(g_1^2+g_2^2)v_{\bar{T}}v_u+\sqrt{2}v_u\Re(T_{\chi_t})
	+\frac{1}{2}v_d v_T\Re(\lambda\Lambda_T^*)-v_d v_s \Re(\lambda\chi_t^*)+4v_{\bar{T}}v_u|\chi_t|^2,\\
	m_{\phi_s,\phi_{\bar{T}}}&=v_s v_{\bar{T}}|\Lambda_T|^2-\frac{1}{\sqrt{2}}v_T\Re(T_\Lambda)
	-v_s v_T\Re(\kappa\Lambda_T^*)-\frac{1}{2}v_d^2\Re(\Lambda_T \chi_d^*)-v_d v_u\Re(\lambda\chi_t^*),\\
	m_{\phi_T,\phi_{\bar{T}}}&=-(g_1^2+g_2^2)v_T v_{\bar{T}}-\frac{1}{\sqrt{2}}v_s \Re(T_\Lambda)
	-\frac{1}{2}v_s^2 \Re(\kappa\Lambda_T^*)+\frac{1}{2}v_d v_u \Re(\lambda\Lambda_T^*)+v_T v_{\bar{T}}|\Lambda_T|^2,\\
	m_{\phi_{\bar{T}},\phi_{\bar{T}}}&=m_{\bar{T}}^2+\frac{1}{4}(g_1^2+g_2^2)(v_d^2-v_u^2-2v_T^2+6v_{\bar{T}})+
	\frac{1}{2}(v_s^2+v_T^2)|\Lambda_T|^2+2v_u^2|\chi_t|^2.
\end{align}
This matrix is diagonalized by $Z^{\rm H}$:
\begin{eqnarray}
	Z^{\rm H}m_h^2 Z^{\rm H,\dagger}=m_{2,h}^{\rm dia}\nonumber
\end{eqnarray}
with
\begin{eqnarray}
	\phi_d=\sum_j Z_{j1}^{\rm H} h_j,~~\phi_u=\sum_j Z_{j2}^{\rm H} h_j,~~\phi_s=\sum_j Z_{j3}^{\rm H} h_j,~~\phi_T=\sum_j Z_{j4}^{\rm H} h_j,~~\phi_{\bar{T}}=\sum_j Z_{j5}^{\rm H}.\nonumber
\end{eqnarray}

In the basis ($\tilde{W}^-$,$\tilde{H}_d^-$,$\tilde{T}^-$), ($\tilde{W}^+$,$\tilde{H}_u^+$,$\tilde{T}^+$), the definition of mass matrix for charginos is given by
\begin{eqnarray}
	m_{\tilde{\chi}^-}=\begin{pmatrix}
		M_2 & \frac{1}{\sqrt{2}}g_2 v_u & g_2 v_T\\
		\frac{1}{\sqrt{2}}g_2 v_d & \frac{1}{\sqrt{2}} v_s \lambda & -v_d \chi_d\\
		g_2 v_{\bar{T}} & -v_u \chi_t & \frac{1}{\sqrt{2}} \Lambda_T v_s
	\end{pmatrix}
\end{eqnarray}

This matrix is diagonalized by U and V
\begin{eqnarray}
	U^*m_{\tilde{\chi}^-}V^{\dagger}=m_{\tilde{\chi}^-}^{dia}\nonumber
\end{eqnarray}
with
\begin{eqnarray}
	\tilde{W}^-=\sum_j U^*_{j1}\lambda_j^-,~~~~\tilde{H}_d^-=\sum_j U^*_{j2}\lambda_j^-,~~~~\tilde{T}^-=\sum_j U^*_{j3}\lambda_j^-\nonumber\\
	\tilde{W}^+=\sum_j V^*_{1j}\lambda_j^+,~~~~\tilde{H}_u^+=\sum_j V^*_{2j}\lambda_j^+,~~~~\tilde{T}^+=\sum_j V^*_{3j}\lambda_j^+\nonumber
\end{eqnarray}
\section{Tadpole equation and some corresponding vertexes}
The CP-even tree level part of tadpole are given by
\begin{align} 
	\frac{\partial V}{\partial \phi_{d}} &= +\frac{1}{8} \Big(g_{1}^{2} + g_{2}^{2}\Big)v_d \Big(2 v_{\bar{T}}^{2}  -2 v_{T}^{2}  - v_{u}^{2}  + v_{d}^{2}\Big)+\frac{1}{4} \Big(v_{\bar{T}} \Big(\Big(-2 v_d v_s \chi_d  + v_T v_u \lambda \Big)\Lambda_T^*\nonumber\\
	&-2 v_s v_u \lambda \chi_t^* \Big)- v_{s}^{2} v_u \lambda \kappa^* +\Big(2 v_d \Big(v_{s}^{2} + v_{u}^{2}\Big)\lambda  + v_u \Big(\Lambda_T v_T v_{\bar{T}}  - v_s \Big(2 \Big(v_{\bar{T}} \chi_t  + v_T \chi_d \Big) + v_s \kappa \Big)\Big)\Big)\lambda^*\nonumber\\
	& +4 v_d \Big(\sqrt{2} v_T {\Re\Big(T_{\chi_d}\Big)}  + m_{H_d}^2\Big)\nonumber \\ 
	&-2 \Big(\sqrt{2} v_s v_u {\Re\Big(T_{\lambda}\Big)}  + \Big(v_d \Big(-2 \Big(2 v_{T}^{2}  + v_{d}^{2}\Big)\chi_d  + \Lambda_T v_s v_{\bar{T}} \Big) + v_s v_T v_u \lambda \Big)\chi_d^* \Big)\Big)\\ 
	\frac{\partial V}{\partial \phi_{u}} &= +\frac{1}{8} \Big(g_{1}^{2} + g_{2}^{2}\Big)v_u \Big(-2 v_{\bar{T}}^{2}  + 2 v_{T}^{2}  - v_{d}^{2}  + v_{u}^{2}\Big)\nonumber \\ 
	&+\frac{1}{4} \Big(\Big(4 v_{u}^{3}  + 8 v_{\bar{T}}^{2} v_u \Big)|\chi_t|^2 +v_T \Big(-2 v_d v_s \lambda \chi_d^*  + \Big(-2 v_s v_u \chi_t  + v_d v_{\bar{T}} \lambda \Big)\Lambda_T^* \Big)\nonumber \\ 
	&+\Big(2 \Big(v_{d}^{2} + v_{s}^{2}\Big)v_u \lambda  + v_d \Big(\Lambda_T v_T v_{\bar{T}}  - v_s \Big(2 \Big(v_{\bar{T}} \chi_t  + v_T \chi_d \Big) + v_s \kappa \Big)\Big)\Big)\lambda^* \nonumber \\ 
	&+v_u \Big(-2 \Lambda_T v_s v_T \chi_t^*  + 4 \Big(\sqrt{2} v_{\bar{T}} {\Re\Big(T_{\chi_t}\Big)}  + m_{H_u}^2\Big)\Big)\nonumber\\
	&+v_d \Big(-2 \sqrt{2} v_s {\Re\Big(T_{\lambda}\Big)}  + \lambda \Big(-2 v_s v_{\bar{T}} \chi_t^*  - v_{s}^{2} \kappa^* \Big)\Big)\Big)\\ 
	\frac{\partial V}{\partial \phi_s} &= \frac{1}{4} \Big(\Big(- v_{d}^{2} v_{\bar{T}} \chi_d  + v_s \Big(2 \Lambda_T \Big(v_{T}^{2} + v_{\bar{T}}^{2}\Big) -2 v_T v_{\bar{T}} \kappa \Big) - v_T v_{u}^{2} \chi_t \Big)\Lambda_T^* +\Big(-2 v_d v_T v_u \lambda  - \Lambda_T v_{d}^{2} v_{\bar{T}} \Big)\chi_d^* \nonumber \\ 
	&+\Big(-2 v_d v_{\bar{T}} v_u \lambda  - \Lambda_T v_T v_{u}^{2} \Big)\chi_t^* +\Big(-2 v_d v_s v_u \lambda  + 4 v_{s}^{3} \kappa \Big)\kappa^* +v_s \Big(-2 \Lambda_T v_T v_{\bar{T}} \kappa^*  + 4 m_S^2 \Big)\nonumber \\ 
	&+2 \Big(- v_d v_u \Big(v_{\bar{T}} \chi_t  + v_s \kappa  + v_T \chi_d \Big) + v_s \Big(v_{d}^{2} + v_{u}^{2}\Big)\lambda \Big)\lambda^*\nonumber\\
	&+\sqrt{2} \Big(-2 v_d v_u {\Re\Big(T_{\lambda}\Big)}  -2 v_T v_{\bar{T}} {\Re\Big(T_{\Lambda_T}\Big)}  + v_{s}^{2} \Big(T_{\kappa}^* + T_{\kappa}\Big)\Big)\Big)\\ 
	\frac{\partial V}{\partial \phi_T} &= +\frac{1}{4} \Big(g_{1}^{2} + g_{2}^{2}\Big)v_T \Big(-2 v_{\bar{T}}^{2}  + 2 v_{T}^{2}  - v_{d}^{2}  + v_{u}^{2}\Big)\nonumber \\ 
	&+\frac{1}{4} \Big(4 m_T^2 v_T +\Big(2 \Lambda_T v_T \Big(v_{s}^{2} + v_{\bar{T}}^{2}\Big) + v_d v_{\bar{T}} v_u \lambda  - v_s \Big(v_s v_{\bar{T}} \kappa  + v_{u}^{2} \chi_t \Big)\Big)\Lambda_T^* \nonumber \\ 
	&+\Lambda_T \Big(- v_{s}^{2} v_{\bar{T}} \kappa^*  - v_s v_{u}^{2} \chi_t^* \Big)+v_d \Big(2 \Big(4 v_d v_T \chi_d  - v_s v_u \lambda \Big)\chi_d^*  + v_u \Big(-2 v_s \chi_d  + \Lambda_T v_{\bar{T}} \Big)\lambda^* \Big)\nonumber \\ 
	&-2 \sqrt{2} v_s v_{\bar{T}} {\Re\Big(T_{\Lambda_T}\Big)} +2 \sqrt{2} v_{d}^{2} {\Re\Big(T_{\chi_d}\Big)} \Big)\\ 
	\frac{\partial V}{\partial \phi_{\bar{T}}} &= +\frac{1}{4} \Big(g_{1}^{2} + g_{2}^{2}\Big)v_{\bar{T}} \Big(2 v_{\bar{T}}^{2}  -2 v_{T}^{2}  - v_{u}^{2}  + v_{d}^{2}\Big)\nonumber \\ 
	&+\frac{1}{4} \Big(4 v_{\bar{T}} \Big(2 v_{u}^{2} |\chi_t|^2  + m_{\bar{T}}^2\Big)+\Big(2 \Lambda_T \Big(v_{s}^{2} + v_{T}^{2}\Big)v_{\bar{T}}  + v_d v_T v_u \lambda  - v_s \Big(v_{d}^{2} \chi_d  + v_s v_T \kappa \Big)\Big)\Lambda_T^* \nonumber \\ 
	&- \Lambda_T v_{s}^{2} v_T \kappa^* +v_d v_u \Big(-2 v_s \chi_t  + \Lambda_T v_T \Big)\lambda^* +v_s \Big(-2 \Big(\sqrt{2} v_T {\Re\Big(T_{\Lambda_T}\Big)}  + v_d v_u \lambda \chi_t^* \Big) - \Lambda_T v_{d}^{2} \chi_d^* \Big)\nonumber \\ 
	&+\sqrt{2} v_{u}^{2} \Big(T_{{\chi,t}^*} + T_{\chi_t}\Big)\Big)
\end{align} 
Then we can identify the $m_{H_d}^2$, $m_{H_u}^2$, $m_S^2$, $m_T^2$ and $m_{\bar{T}}^2$ by the minimum conditions of the scalar potential.

Here, we show some corresponding vertexes in this model. Their concrete forms are shown
as
\begin{eqnarray}
	C^L_{Z\chi^+_i\chi^-_j}=&&\frac{1}{2}\big(-2g_1\sin\theta_WU^*_{j3}U_{i3}+2g_2\cos\theta_WU^*_{j1}U_{i1}+(-g_1\sin\theta_W+g_2\cos\theta_W)U^*_{j2}U_{i2}\big)\\
	C^R_{Z\chi^+_i\chi^-_j}=&&\frac{1}{2}\big(-2g_1\sin\theta_WV^*_{j3}V_{i3}+2g_2\cos\theta_WV^*_{j1}V_{i1}+(-g_1\sin\theta_W+g_2\cos\theta_W)V^*_{j2}V_{i2}\big)\\
	C^L_{Z\chi^{++}\chi^{--}}=&&C^R_{h\chi^{++}\chi^{--}}=\big(-g_1\sin\theta_W+g_2\cos\theta_W\big)\\
	C^L_{h_k\chi^+_i\chi^-_j}=&&-\frac{1}{2}\big[g_2U^*_{j1}(2V^*_{i3}Z^H_{k4}+\sqrt{2}V^*_{i2}Z^H_{k2})+U^*_{j2}(-2\chi_d V^*_{i3}Z^H_{k1}+\sqrt{2}g_2V^*_{i1}Z^H_{k3}+\sqrt{2}\lambda V^*_{i2}Z^H_{k3})\nonumber\\ 
	&&+U^*_{j3}(-2\chi_tV^*_{i2}Z^H_{k2}+2g_2V^*_{i1}Z^H_{k5}+\sqrt{2}\Lambda_TV^*_{i3}Z^H_{k3})\big]\\
	C^R_{h_k\chi^+_i\chi^-_j}=&&-\frac{1}{2}\big[g_2U_{i1}(2V_{j3}Z^H_{k4}+\sqrt{2}V_{j2}Z^H_{k2})+U_{i2}(-2\chi_d^* V_{j3}Z^H_{k1}+\sqrt{2}g_2V_{j1}Z^H_{k3}+\sqrt{2}\lambda^* V_{j2}Z^H_{k3})\nonumber\\ 
	&&+U_{i3}(-2\chi_t^*V_{j2}Z^H_{k2}+2g_2V_{j1}Z^H_{k5}+\sqrt{2}\Lambda_T^*V_{j3}Z^H_{k3})\big]\\
	C^L_{h_k\chi^{++}\chi^{--}}=&&\frac{1}{\sqrt{2}}\Lambda_T Z^H_{k3}\\
	C^R_{h_k\chi^{++}\chi^{--}}=&&\frac{1}{\sqrt{2}}\Lambda_T^* Z^H_{k3}
\end{eqnarray}

For the coupling between Higgs particles and charged scalar particles is too complicated, we calculate it by computer.
\begin{eqnarray}
	V_{\rm scalar}=&&W^*_iW^i+\frac{1}{2}\sum_{a}g_2^2\big[H_u^\dagger\frac{\sigma^a}{2}H_u+H_d^\dagger\frac{\sigma^a}{2}H_d+Tr(T^\dagger\sigma^a T)+Tr(\bar{T}^\dagger\sigma^a \bar{T})\big]^2\nonumber\\
	&&g^2_1\big[\frac{1}{2}H_u^\dagger H_u-\frac{1}{2}H_d^\dagger H_d+T^\dagger T-\bar{T}^\dagger\bar{T}+\frac{1}{6}\tilde{Q}^\dagger\tilde{Q}-\frac{1}{2}\tilde{L}^\dagger\tilde{L}\nonumber\\
	&&-\frac{2}{3}\tilde{u}^*_R\tilde{u}_R+\frac{1}{3}\tilde{d}^*_R\tilde{d}_R+\tilde{e}^*_R\tilde{e}_R+Tr(T^\dagger T)-Tr(\tilde{T}^\dagger\tilde{T})\big]^2+V_{soft}\nonumber
\end{eqnarray}
where $W_i=\frac{\partial W}{\partial\phi_i}$.
And the coefficient C are as follow:
\begin{eqnarray}
	C_{h_k\phi_i\phi_j}=\frac{\partial^3 V_{\rm scalar}}{\partial h_k\partial \phi_i\partial\phi_j}\big|_{<H^0_{d,u}>=\frac{v_{d,u}}{\sqrt{2}},<S>=\frac{v_s}{\sqrt{2}},<T>=\frac{v_T}{\sqrt{2}},<\bar{T}>=\frac{v_{\bar T}}{\sqrt{2}},<\rm the~other ~fields>=0}\nonumber\\
\end{eqnarray} 
where $\phi_i$ represent the scalar field: $H^{\pm}$, $H^{\pm\pm}$, $\tilde{u}_L$, $\tilde{u}_R$, $\tilde{d}_L$, $\tilde{d}_R$, $\tilde{e}_L$ and $\tilde{e}_R$.


\begin{thebibliography}{99}
\bibitem{mh-ATLAS}G.~Aad  et al. (ATLAS Collaboration),  \emph{Phys.~Lett.} {\bf B 716} (2012) 1.
\bibitem{mh-CMS}S.~Chatrchyan et al. (CMS Collaboration),  \emph{Phys.~Lett.} {\bf B 716} (2012) 30.
\bibitem{PDG}
R.~L.~Workman \textit{et al.} [Particle Data Group],
PTEP \textbf{2022}, 083C01 (2022)
doi:10.1093/ptep/ptac097

\bibitem{Weinberg:1975gm}
S.~Weinberg,
Phys. Rev. D \textbf{13}, 974-996 (1976)
doi:10.1103/PhysRevD.19.1277

\bibitem{Susskind:1978ms}
L.~Susskind,
Phys. Rev. D \textbf{20}, 2619-2625 (1979)
doi:10.1103/PhysRevD.20.2619

\bibitem{Ellwanger:2006rm}
U.~Ellwanger and C.~Hugonie,
Mod. Phys. Lett. A \textbf{22}, 1581-1590 (2007)
doi:10.1142/S0217732307023870
[arXiv:hep-ph/0612133 [hep-ph]].

\bibitem{Ananthanarayan:1995zr}
B.~Ananthanarayan and P.~N.~Pandita,
Phys. Lett. B \textbf{371}, 245-251 (1996)
doi:10.1016/0370-2693(96)00010-X
[arXiv:hep-ph/9511415 [hep-ph]].

\bibitem{Ananthanarayan:1996zv}
B.~Ananthanarayan and P.~N.~Pandita,
Int. J. Mod. Phys. A \textbf{12}, 2321-2342 (1997)
doi:10.1142/S0217751X97001353
[arXiv:hep-ph/9601372 [hep-ph]].

\bibitem{Mason:2009iq}
J.~D.~Mason,
Phys. Rev. D \textbf{80}, 015026 (2009)
doi:10.1103/PhysRevD.80.015026
[arXiv:0904.4485 [hep-ph]].

\bibitem{TNMSSM}
K.~Agashe, A.~Azatov, A.~Katz and D.~Kim,
Phys. Rev. D \textbf{84}, 115024 (2011)
doi:10.1103/PhysRevD.84.115024
[arXiv:1109.2842 [hep-ph]].

\bibitem{Espinosa:1991gr}
J.~R.~Espinosa and M.~Quiros,
Phys. Lett. B \textbf{279}, 92-97 (1992)
doi:10.1016/0370-2693(92)91846-2

\bibitem{Espinosa:1992hp}
J.~R.~Espinosa and M.~Quiros,
Phys. Lett. B \textbf{302}, 51-58 (1993)
doi:10.1016/0370-2693(93)90634-T
[arXiv:hep-ph/9212305 [hep-ph]].

\bibitem{Espinosa:1998re}
J.~R.~Espinosa and M.~Quiros,
Phys. Rev. Lett. \textbf{81}, 516-519 (1998)
doi:10.1103/PhysRevLett.81.516
[arXiv:hep-ph/9804235 [hep-ph]].






\bibitem{hMZ}
M.~Gonzalez-Alonso and G.~Isidori,
Phys. Lett. B \textbf{733}, 359-365 (2014)
doi:10.1016/j.physletb.2014.05.004
[arXiv:1403.2648 [hep-ph]].

\bibitem{hMZ1}
G.~Isidori, A.~V.~Manohar and M.~Trott,
Phys. Lett. B \textbf{728}, 131-135 (2014)
doi:10.1016/j.physletb.2013.11.054
[arXiv:1305.0663 [hep-ph]].

\bibitem{hMZ2}
S.~Alte, M.~K\"onig and M.~Neubert,
JHEP \textbf{12}, 037 (2016)
doi:10.1007/JHEP12(2016)037
[arXiv:1609.06310 [hep-ph]].

\bibitem{Zhao:hmz}
S.~M.~Zhao, T.~F.~Feng, J.~B.~Chen, J.~J.~Feng, G.~Z.~Ning and H.~B.~Zhang,
Phys. Rev. D \textbf{97}, no.9, 095043 (2018)
doi:10.1103/PhysRevD.97.095043
[arXiv:1805.05048 [hep-ph]].

\bibitem{hMZ4}
A.~L.~Kagan, G.~Perez, F.~Petriello, Y.~Soreq, S.~Stoynev and J.~Zupan,
Phys. Rev. Lett. \textbf{114}, no.10, 101802 (2015)
doi:10.1103/PhysRevLett.114.101802
[arXiv:1406.1722 [hep-ph]].

\bibitem{hMZ5}
G.~T.~Bodwin, H.~S.~Chung, J.~H.~Ee, J.~Lee and F.~Petriello,
Phys. Rev. D \textbf{90}, no.11, 113010 (2014)
doi:10.1103/PhysRevD.90.113010
[arXiv:1407.6695 [hep-ph]].

\bibitem{ATLAS:hMZ}
G.~Aad \textit{et al.} [ATLAS],
JHEP \textbf{10}, 013 (2021)
doi:10.1007/JHEP10(2021)013
[arXiv:2104.13240 [hep-ex]].

\bibitem{ATLAS:2023yqk}
G.~Aad \textit{et al.} [ATLAS and CMS],
	Phys. Rev. Lett. \textbf{132}, no.2, 021803 (2024)
	doi:10.1103/PhysRevLett.132.021803
	[arXiv:2309.03501 [hep-ex]].

\bibitem{cgz}
L.~Bergstrom and G.~Hulth,
Nucl. Phys. B \textbf{259}, 137-155 (1985)
[erratum: Nucl. Phys. B \textbf{276}, 744-744 (1986)]
doi:10.1016/0550-3213(85)90302-5


\bibitem{QCD}
M.~Spira, A.~Djouadi and P.~M.~Zerwas,
Phys. Lett. B \textbf{276}, 350-353 (1992)
doi:10.1016/0370-2693(92)90331-W

\bibitem{direct}
V.~L.~Chernyak and A.~R.~Zhitnitsky,
Nucl. Phys. B \textbf{201}, 492 (1982)
[erratum: Nucl. Phys. B \textbf{214}, 547 (1983)]
doi:10.1016/0550-3213(83)90251-1

\bibitem{direct1}
N.~H.~Fuchs and M.~D.~Scadron,
Phys. Rev. D \textbf{20}, 2421 (1979)
doi:10.1103/PhysRevD.20.2421

\bibitem{direct2}
M.~Beneke, G.~Buchalla, M.~Neubert and C.~T.~Sachrajda,
Nucl. Phys. B \textbf{591}, 313-418 (2000)
doi:10.1016/S0550-3213(00)00559-9
[arXiv:hep-ph/0006124 [hep-ph]].

\bibitem{higgs decay1}
A.~Djouadi,
Phys. Rept. \textbf{457}, 1-216 (2008)
doi:10.1016/j.physrep.2007.10.004
[arXiv:hep-ph/0503172 [hep-ph]].

\bibitem{higgs decay2}
A.~Djouadi,
Phys. Rept. \textbf{459}, 1-241 (2008)
doi:10.1016/j.physrep.2007.10.005
[arXiv:hep-ph/0503173 [hep-ph]].


\end{thebibliography}
\end{document}